\documentclass[draft,nofootinbib,showpacs]{revtex4}    

\usepackage{psfig}
\usepackage{colordvi}
\usepackage{graphics}
\usepackage{lscape}
\usepackage{rotfloat}
\usepackage{rotating}
\usepackage{amsmath}
\usepackage{amssymb}
\usepackage{graphicx}

\usepackage{flexisym}
\usepackage{breqn}
\newcommand{\bdm}{\begin{dmath}}
\newcommand{\edm}{\end{dmath}}
\newcommand{\bdms}{\begin{dmath*}}
\newcommand{\edms}{\end{dmath*}}
\newcommand{\bdg}{\begin{dgroup*}}
\newcommand{\edg}{\end{dgroup*}}
\def\lsim{\mathrel{\rlap{\lower4pt\hbox{\hskip1pt$\sim$}}
    \raise1pt\hbox{$<$}}}                % less than or approx. symbol
\def\gsim{\mathrel{\rlap{\lower4pt\hbox{\hskip1pt$\sim$}}
    \raise1pt\hbox{$>$}}}                % greater than or approx. symbol

\newcommand{\tendto}{\mathop{\longrightarrow}}

\def\smn{{\sigma_{\mu\nu}}}

\def\openone{\leavevmode\hbox{\small1\kern-3.3pt\normalsize1}}

\def\Zq{Z_{\rm q}}
\def\Zm{Z_{\rm m}}

\def\ZS{Z_{\rm S}}
\def\ZP{Z_{\rm P}}
\def\ZV{Z_{\rm V}}
\def\ZA{Z_{\rm A}}
\def\ZT{Z_{\rm T}}

\newcommand{\be}{\begin{equation}}
\newcommand{\ee}{\end{equation}}
\newcommand{\bea}{\begin{eqnarray}} 
\newcommand{\eea}{\end{eqnarray}}

\newcommand{\pslash}{{\not{\hspace{-0.001cm}p}}}  
\newcommand{\qslash}{{\not{\hspace{-0.001cm}q}}}

\newcommand{\ggcf}{\frac{g^2 C_F}{16 \, \pi^2}\; }  
\begin{document}

\title{Perturbative renormalization functions of local operators \\
for staggered fermions with stout improvement}

\author{M. Constantinou, M. Costa and H. Panagopoulos\,~}

\vskip 0.25cm
\email{marthac@ucy.ac.cy,kosta.marios@ucy.ac.cy, haris@ucy.ac.cy}

\vskip 0.25cm
\affiliation{
$\,$ \\
$Department\,\, of\,\, Physics,\,\, University\,\, of\,\,Cyprus,$\\
$POB\,\, 20537,\,\, 1678\,\, Nicosia,\,\, Cyprus$}

%%%%%%%%%%%%%%%%%%%%%%%%%%%%%%%%%%%%%%%%%%%%%%%%%%%%%%%%%%%%%%%%%%%%%%%%%%%%% 

\begin{abstract}
In this paper we present the perturbative computation of the
renormalization functions for the quark field and for a complete set
of ultralocal fermion bilinears. The computation of the relevant Green's functions are
carried out at one-loop level for the staggered action using massive
fermions. The gluon links which appear both in the fermion action and
in the definition of the bilinears are improved by applying a
stout smearing procedure up to two times, iteratively. In the gluon sector we
employ the Symanzik improved gauge action for different sets of
values of the Symanzik coefficients. The renormalization functions are presented in (two
variants of) the RI$'$ and in the $\overline{\rm MS}$ renormalization
scheme; the dependence on all stout parameters, as well as on the
fermion mass, the gauge fixing parameter and the renormalization
scale, is shown explicitly. This work is related to our recent paper
[Phys. Rev. D 86, 094512 (2012)]. To make our results
easily accessible to the reader, we include them in the distribution
package of this paper as a Mathematica input file, ``Staggered.m''\,.

\end{abstract}


\maketitle

\section{Introduction}

In recent years, significant improvements have been made in the use of
matrix elements of operators made out of quark fields to extract
mass spectra, decay constants, and a plethora of hadronic properties
\cite{Hagler:2009ni,Doi:2012ab,Lin:2012ev}. Although naive
(unimproved) staggered fermions were introduced more than three decades
ago~\cite{Kogut:1974ag}, their discretization errors and
their relatively large taste mixing posed a limit on the accuracy of
results from simulations, despite their relatively low computational
cost. This situation called for improvement; the outcome of such
efforts was some of the most accurate discretizations used to date for
high-precision simulations. One specific direction regards improving
the fermion action (see, e.g. \cite{Bazavov:2012zad,Allison:2008xk});
in particular, the introduction of stout links in the action which has
recently been put to use~\cite{Bali:2011qj,Bali:2012jv} allows
simulations to be carried out at near physical parameters. Compared to
most other improved formulations of staggered fermions, the above
action, as well as the HISQ action, lead to smaller taste violating
effects~\cite{Aoki:2005vt,Borsanyi:2011bm,Bazavov:2011nk}. 

Changes in the lattice action and in the discretization of operators
imply that renormalization functions must be determined afresh, either
perturbatively or nonperturbatively. In many cases nonperturbative
estimates of renormalization functions are very difficult to obtain,
due to complications such as possible mixing with operators of equal
or lower dimension, whose signals are hard to disentangle. For this
reason, and in order to provide cross-estimates which have a reduced
systematic error, the perturbative study of a variety of fermion
operators is widely employed in numerical simulations of QCD on the
lattice (see, e.g. \cite{Capitani:2002mp} and references therein, also
\cite{Patel:1992vu,Aoki:2002iq,Mason:2005bj,Skouroupathis:2007jd,Alexandrou:2010me}).

Within the staggered formulation using massive fermions we compute
the fermion propagator and Green's functions of a set of local
taste-singlet bilinears $\cal O$ [scalar (S), pseudoscalar (P), vector
(V), axial (A), and tensor (T)]. Our computation is performed to one loop
and to lowest order in the lattice spacing, $a$. We also extract from
the above the renormalization functions (RFs) of the quark field $Z_q$,
quark mass $Z_m$\,, and fermion bilinears $Z_{\cal O}$.
This work is a continuation to our recent paper~\cite{Bali:2012jv},
in which we presented our perturbative results for $Z_q,\,\ZT$\,, and
$\ZS$; it is the first one-loop computation of these quantities using
staggered fermions with stout links. In the present paper, we provide
the details of the perturbative calculation and our results for the
propagator and for the Green's functions, as well as the
renormalization functions of all operators, including the vector,
axial, and pseudoscalar cases. Older results with staggered
fermions~\cite{Patel:1992vu} in the absence of stout smearing and for
the Wilson gluon action are in complete agreement with our results;
perturbartive results related to alternative improvements of the
staggered action can be found, e.g., in Refs.~\cite{Lee:2002ui,Kim:2010fj}.

Stout links~\cite{Morningstar:2003gk} rather than ordinary links
have been used both in the fermion action and in bilinear
operators. Following Ref.~\cite{Bali:2012jv}, we use two steps of stout
smearing with generic smearing parameters ($\omega_1$, $\omega_2$). We
emphasize that the results for the bilinear Green's functions depend on four
stout parameters, two due to the action smearing ($\omega_{A_1},\,
\omega_{A_2}$) and two more coming from the smearing of the operator
($\omega_{{\cal O}_1},\,\omega_{{\cal O}_2}$); no numerical value needs to be
specified for these parameters. The extension to further steps of
stout smearing can be achieved with relative ease. For gluons we
employ the Symanzik improved action. Our final expressions for the
Green's functions exhibit a rather nontrivial dependence on the
external momentum ($q$) and the fermion mass ($m$), and they are
polynomial functions of the gauge parameter ($\alpha$), stout parameters
($\omega_{A_i},\,\omega_{{\cal O}_i}$), and coupling constant ($g$);
furthermore, most numerical coefficients in these expressions depend
on the Symanzik parameters of the gluon action.

The one-loop expressions for the renormalization functions are presented
in the mass-independent RI$'$ scheme; for the vector and axial
renormalization functions we also employ an alternative RI$'$ scheme
which might be more useful in renormalizing nonperturbative matrix
elements. Furthermore, for comparison with experimental determinations
and phenomenological estimates, it is useful to present our results
also in the $\overline{\rm MS}$ scheme; we do so, paying particular
attention to the possible alternative definitions of $\gamma_5$.

Results for $\Zq,\,\Zm,\,Z_{\cal O}$ exist for simpler actions to
${\cal O}(g^4)$ and/or ${\cal O}(g^2 a^n)$, see e.g., 
Refs.~\cite{Skouroupathis:2007jd,Skouroupathis:2008mf}
for two-loop renormalization of flavor singlet and nonsinglet local
fermion bilinears, Ref.~\cite{Mason:2005bj} for $\Zm$ to two loops,
Ref.~\cite{Capitani:2000xi} for one-loop renormalization of the fermion
propagator and bilinears to ${\cal O}(a^1)$, and
Refs.~\cite{Constantinou:2009tr,Alexandrou:2010me,Alexandrou:2012mt}
for the fermion propagator and bilinears with 0 and 1 derivatives to
one-loop and to ${\cal O}(a^2)$. The extension of the present
computation beyond one loop and/or beyond ${\cal O}(a^0)$ becomes
exceedingly complicated: One reason for this is the appearance of
divergences in nontrivial corners of the Brillouin zone; also, a
two-loop calculation requires vertices with up to four gluons, which are
extremely lengthy in the presence of stout links (estimated length:
$>10^6$ terms).

To make our results easily accessible, we accompany this paper with an
electronic document in the form of a Mathematica input file
(``Staggered.m''), allowing the reader to extract the expressions for
many choices of the action parameters. This document contains 
  the one-loop inverse fermion propagator,
  the one-loop amputated Green's functions relevant to all
  ultralocal operators, and the renormalization functions in
  the RI$'$ scheme for the fermion field and for all bilinears. In
  addition, in Staggered.m we provide the expressions for the one-
  and two-gluon ``doubly stout'' links for different stout parameters in
  the first and second smearing step.

The outline of this paper is as follows: Sec. II regards a brief
theoretical background in which we introduce the formulation of the
action and of the operators which we employ. Section III contains a
summary of the calculational procedure for the fermion propagator and
for the Green's functions of the bilinear operators. We also present
the most general results for these quantities using the tree-level
Symanzik improved gluon action. The renormalization functions are
derived in Sec. IV for different renormalization schemes, and we
provide their expressions for tree-level Symanzik gluons. Finally, we
conclude in Sec. V with a discussion of our results and possible
future extensions of our work. For completeness, we have included 3
Appendices containing: A: the stout smeared links, B: the numerical
results of the propagator for the Wilson and tree-level Symanzik
actions, and C: a description of the Mathematica file ``Staggered.m''\,.

%For completeness, we have included 
%Appendix A: Stout Smeared Links,
%Appendix B: Numerical Results of the Propagator, and 
%Appendix C: Notation in Mathematica File Staggered.m\,.

\section{Formulation}

\subsection{Lattice actions}

Our perturbative calculation makes use of the staggered fermion
action. Let us briefly go over the derivation of the latter, starting from the naive fermion action
\be
S^F = a^4\,\sum_{x,\,f,\,\mu}\,\bar{
 \psi}^{f}(x) \left(\gamma_\mu D_\mu\right)\psi^f(x) 
+ a^4\,\sum_{x,\,f}\,m_f\,\bar{\psi}^{f}(x)\,\psi^f(x) \,,
\label{naive_action}
\ee
where $f$ is a flavor index. Given that, to one loop in perturbation
theory, the quantities of interest do not depend on the number of
flavors, we will drop the index $f$ from this point on. The covariant
derivative $D_\mu$ is defined as
\be
D_\mu \psi(x)=\frac{1}{2a}\Big[U_\mu (x) \psi (x + a \hat{\mu})
- U_\mu^\dagger (x - a \hat{\mu}) \psi (x - a \hat{\mu})\Big]\,.
\ee
The absence of the Wilson term in the naive action of
Eq.~(\ref{naive_action}) leads to the well-known doubling
problem. The standard passage to the staggered action entails the
following change of basis:
\bea
\psi(x) = \gamma_x\,\chi(x),\quad 
\bar\psi(x) = \bar\chi(x)\,\gamma_x^\dagger,\nonumber \\[2ex]
 \gamma_x=\gamma_1^{n_1}\,\gamma_2^{n_2}\,\gamma_3^{n_3}\,\gamma_4^{n_4},
\quad x =(a\,n_1,a\,n_2,a\,n_3,a\,n_4),\quad n_i\,\, \epsilon\,\, {\mathbb Z}\,. \label{transf}
\eea
Using the equalities
\be
\gamma_\mu\,\gamma_x = \eta_\mu(x)\gamma_{x +a\,\hat{\mu}}\quad {\rm and}\quad
\gamma_x^\dagger\,\gamma_x = \openone\,,\qquad \eta_\mu (x)  =
(-1)^{\sum_{\nu < \mu} n_\nu}\,,
\label{equalities}
\ee
the lattice fermion action takes the form
\be
S^F = a^4 \sum_{x} \sum_{\mu} \frac{1}{2a} \, \overline{\chi} (x) 
\eta_\mu (x) \Big[ U_\mu (x) \chi (x + a \hat{\mu})
- U_\mu^\dagger (x - a \hat{\mu}) \chi (x - a \hat{\mu}) \Big] 
+ a^4 \sum_{x} m \overline{\chi} (x) \chi (x)\,.
\label{SFaction1}
\ee
Thus far, we have rewritten the usual lattice action. But the crucial step now 
is that the Dirac matrices have disappeared, and they have been replaced 
by the phase factors $\eta_\mu (x)$; in the new basis, the naive
action consists of four identical parts, one for each value of the spinor
index carried by the spinor $\chi$. Dropping this index altogether
leads to the standard staggered fermion action $S_{\rm stag}$
containing four rather than 16 fermion ``tastes" (doublers).
These four tastes contain a total of 16 components, which are
split over a unit hypercube by assigning only a single fermion
field component to each lattice site.
 
\bigskip
Following the nonperturbative work of Ref.~\cite{Bali:2012jv} we
apply stout smearing according to Ref.~\cite{Morningstar:2003gk} to
all links appearing in $S_{\rm stag}$: Each link 
$U_\mu(x)=exp(i\,g\,a\,A_\mu(x+a\hat{\mu}/2))$ is replaced by
a stout link $\widetilde{U}_\mu(x)$ defined as~\cite{Horsley:2008ap}
\be
\widetilde{U}_\mu(x) = e^{i\,Q_\mu(x)}\, U_\mu(x)\,,
\ee 
where the definition of $Q_\mu(x)$ is
\be
  Q_\mu(x)=\frac{\omega}{2\,i} \left[V_\mu(x) U_\mu^\dagger(x) -
  U_\mu(x)V_\mu^\dagger(x) -\frac{1}{3} {\rm Tr} \,\Big(V_\mu(x)
  U_\mu^\dagger(x) -  U_\mu(x)V_\mu^\dagger(x)\Big)\right] \, .
\label{Q_def}
\ee
$\omega$ is a tunable parameter, called a stout smearing parameter,
and $V_\mu(x)$ represents the sum over all staples associated with the
link, $U_\mu(x)$. In the present work we need the contributions of
$Q_\mu(x)$ up to two gluons, to which the trace terms in Eq.~(\ref{Q_def}) are irrelevant;
the contributions can be read from the terms\\
\bea 
Q^{\rm up\,to\,2{-}gluons}_\mu(x)= \frac{\omega}{2i} \sum_{\rho=\pm
  1}^{\pm 4}\hspace{-0.15cm}&\Bigg(&\hspace{-0.15cm}
        U_\rho(x) U_\mu(x{+}a \hat{\rho}) U^\dagger_\rho(x{+}a
        \hat{\mu}) U^\dagger_\mu(x)
      - U_\mu(x) U_\rho(x{+}a \hat{\mu}) U^\dagger_\mu(x{+}a
      \hat{\rho})
        U^\dagger_\rho(x) \Bigg)
\eea 
($U_{-\rho}(y)\equiv U_\rho^\dagger(y-a \hat{\rho}),\,\,\rho>0$).
The above procedure can be performed iteratively by dressing the
links more than once in order to improve the convergence to the
continuum limit. In the framework of our calculation we use
``doubly stout" links
\be
\widetilde{\widetilde{U}}_\mu(x) = e^{i\,\widetilde{Q}_\mu(x)}\,\widetilde{U}_\mu(x)\,,
\ee
where $\widetilde{Q}$ is defined as in Eq.~(\ref{Q_def}) but using
$\widetilde{U}$ as links (also in the construction of $V_\mu$). Such
links have been employed in numerical simulations in
Refs.~\cite{Bali:2011qj,Borsanyi:2011bm}. To obtain results that are
as general as possible, we use different stout parameters, $\omega$,
in the first ($\omega_1$) and the second ($\omega_2$) smearing
iteration. This allows for further optimization of improvement, by
separate tuning of the two parameters; it also provides a check of the
perturbative calculation by comparing the limit $\omega_1=0$ (or
$\omega_2=0$) to the case of a single step of stout smearing. We smear
both the links in $S_{\rm stag}$ and those in bilinear operators (see
following subsection), so that we have a total of four stout parameters
that we keep different from one another. In Appendix~\ref{appA} we
present the one-gluon link, $U^{(1)}$, for general $\omega_1$ and
$\omega_2$, as well as the two-gluon link, $U^{(2)}$; due to space
limitations, the lengthy expression for $U^{(2)}$ (a total of
$\sim$500 terms) has been presented only for $\omega_2=0$. The general
expression for the two-gluon link  with $0 \neq \omega_1 \neq \omega_2
\neq 0$ is provided in the Mathematica file.

%%%%%%%%%%%%%%%%%%%%%%%%%%%%%%%%%%%%%%%%%%%%%%%%%%%%%%%%%%%%%%%%%%%%%%%%%%%%%%%%%%%%%%%%

\begin{center}
\begin{figure}[!h]
\centerline{\psfig{figure=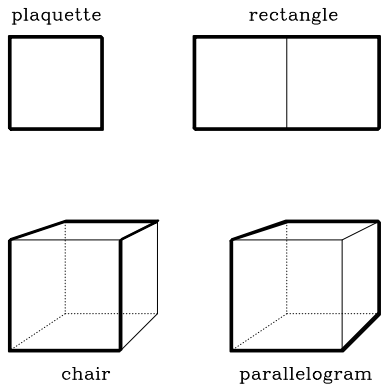,height=5truecm,angle=-90}}
\begin{minipage}{10cm}
\caption{The four Wilson loops of the gluon action.}
\label{figSym}
\end{minipage}
\end{figure}
\end{center}

\bigskip
For gluons we employ the Symanzik improved action, involving
Wilson loops with four and six links ($1\times 1$ {\em plaquette} (a),
$1\times 2$ {\em rectangle} (b), $1\times 2$ {\em chair} (c), and $1\times
1\times 1$ {\em parallelogram} (d) wrapped around an elementary 3D
cube), as shown in Fig.~\ref{figSym}:
\bea
\hspace{-1cm}
S_G=\frac{2}{g_0^2} \Bigl[ &c_0& \sum_{\rm plaq.} {\rm Re\,Tr\,}\{1-U_{\rm plaq.}\}
%\nonumber \\ 
\,+\, c_1 \sum_{\rm rect.} {\rm Re \, Tr\,}\{1- U_{\rm rect.}\} 
\nonumber \\ 
+ &c_2& \sum_{\rm chair} {\rm Re\, Tr\,}\{1-U_{\rm chair}\} 
%\nonumber \\ 
\,+\, c_3 \sum_{\rm paral.} {\rm Re \,Tr\,}\{1-U_{\rm paral.}\}\Bigr]\,.
\label{Symanzik}
\eea
The coefficients $c_i$ can in principle be chosen arbitrarily, subject
to the following normalization condition, which ensures the correct
classical continuum limit of the action
\be
c_0 + 8 c_1 + 16 c_2 + 8 c_3 = 1\,.
\label{norm}
\ee

Some popular choices of values for $c_i$ used in numerical simulations
will be considered in this work and are itemized in Table I
of Ref.~\cite{Alexandrou:2012mt}. They include the Wilson case
($c_0=1$, $c_1=c_2=c_3=0$), and the tree-level Symanzik, TILW (tadpole
improved L\"uscher-Weisz), Iwasaki, and DBW2 (doubly blocked Wilson action) actions. 
In the results presented in paper form we use the tree-level Symanzik action
($c_0=5/3$, $c_1=-1/12$, $c_2=c_3=0$). Our one-loop Feynman diagrams do
not involve pure gluon vertices, and the gluon propagator depends only
on three combinations of the Symanzik parameters: 
\be
c_0 + 8 c_1 + 16 c_2 + 8 c_3 \,=1,\quad c_2 + c_3, \quad c_1 - c_2 - c_3\,.\nonumber
\ee
Therefore, with no loss of generality we choose $c_2=0$.
\bigskip

\subsection{Definition of operators}

In the staggered formalism one defines fields that live on the
corners of four-dimensional elementary hypercubes of the
lattice~\cite{Daniel:1987aa, Patel:1992vu, Ishizuka:1993fs}. The
position of a hypercube inside the lattice is denoted by the index
$y$, where $y$ is a four-vector with components $y_\mu$, which are
even integers ($y_\mu\,\epsilon\,\,2{\mathbb Z}$). The position of a fermion
field component within a specific hypercube is defined by one
additional four-vector index, $C$ ($C_\mu \in \{0,1\}$).

To be able to obtain the correct continuum limit, both for the action
and for operators containing fermions, we relate $\chi$ with the
physical field $Q_{\beta,b}$ ($\beta$: Dirac index, $b$: taste
index). In standard notation:
\be
\chi(y)_C\equiv\chi(ay+aC)/4 =
\sum_{\beta,b}\left(\frac{1}{2}\xi_C\right)_{\beta,b}\,Q_{\beta,b}(y)\,,\quad
Q_{\beta,b}(y) \equiv \frac{1}{2}\,\sum_C\left(\gamma_C\right)_{\beta,b}\,\chi(y)_C\,,
\ee
where $\xi_C$ is defined similarly to $\gamma_C$ [Eq.~(\ref{transf})], that is  
$\xi_C=\xi_1^{C_1}\,\xi_2^{C_2}\,\xi_3^{C_3}\,\xi_4^{C_4},\quad
\xi_\mu=\left(\gamma_\mu^\star \right).$ In terms of the field $Q$ one
can now define fermion bilinear operators as follows:
\be
{\cal O}_{\Gamma,\xi}=\bar Q\,\left(\Gamma\otimes\xi\right)\,Q\,,
\ee
where $\Gamma$ and $\xi$ are arbitrary $4\times4$ matrices acting on
the Dirac and taste indices of $Q_{\beta,b}$, respectively. After
rotating into the staggered basis, the operator ${\cal O}_{\Gamma,\xi}$ 
can be written as~\cite{Patel:1992vu}
\bea
{\cal O}_{\Gamma,\xi} = \sum_{C,D}
\bar\chi(y)_C\,\left(\overline{\Gamma
  \otimes\xi}\right)_{CD}\,\chi(y)_D\,,
\label{O_general}\\
\left(\overline{\Gamma\otimes\xi}\right)_{CD} \equiv
\frac{1}{4}\,{\rm
  Tr}\left[\gamma^\dagger_C\,\Gamma\,\gamma_D\,\xi\right]\,.
\label{gamma1}
\eea
In this work we focus on taste-singlet operators, thus $\xi=\openone$.

The operator of Eq.~(\ref{O_general}) is clearly not gauge
invariant, since $\bar\chi$ and $\chi$ are defined at
different points of the hypercube. To restore gauge
invariance, we insert the average of products of gauge link variables
along all possible shortest paths connecting the sites $y+C$ and
$y+D$. This average is denoted by $U_{C,D}$ and the gauge invariant-operator is now
\be
{\cal O}_\Gamma \equiv {\cal O}_{\Gamma,\openone} = \sum_{C,D} \bar\chi(y)_C\,
\left(\overline{\Gamma\otimes\openone}\right)_{CD}\,
U_{C,D}\,\chi(y)_D\,.
\label{Oper}
\ee
From the definition of Eq.~(\ref{gamma1}), as well as the
equalities of Eq.~(\ref{equalities}), we can further simplify the
expression for the operator ${\cal O}_\Gamma$, using
\bea
\frac{1}{4}{\rm Tr}\left[\gamma_C^\dagger\,\openone\,\gamma_D\right]
&=& \delta_{C,D}\,,\nonumber\\
\frac{1}{4}\,{\rm Tr}\left[\gamma^\dagger_C\,\gamma_\mu\,\gamma_D\,\right]&=&
\delta_{C,D+\hat{\mu}}\,\,\eta_\mu(D)\,,\nonumber\\
\label{Oper2}
\frac{1}{4}{\rm Tr}\left[\gamma_C^\dagger\,\smn\,\gamma_D\right]
&=&\frac{1}{i}\,\delta_{C,D+\hat{\mu}+\hat{\nu}}\,\,\eta_\nu(D)\,\,\eta_\mu(D+\hat{\nu})\,,\nonumber\\
\frac{1}{4}\,{\rm Tr}\left[\gamma^\dagger_C\,\gamma_5\,\gamma_\mu\,\gamma_D\,\right]&=& \delta_{C,D+\hat{\mu}+(1,1,1,1)}\,\,\eta_\mu(D)\,\eta_1(D+\hat{\mu})\,\eta_2(D+\hat{\mu})\,\eta_3(D+\hat{\mu})\,\eta_4(D+\hat{\mu})\,,\nonumber\\
\frac{1}{4}\,{\rm Tr}\left[\gamma^\dagger_C\,\gamma_5\,\gamma_D\,\right]&=&
\delta_{C,D+(1,1,1,1)}\,\,\eta_1(D)\,\eta_2(D)\,\eta_3(D)\,\eta_4(D)\,,
\eea
where $\sigma_{\mu\,\nu}=[\gamma_\mu,\gamma_\nu]/(2i)$.
Here and below, in expressions such as $D+\hat{\mu}$, the sum is to be
taken modulo 2. Using Eq.~(\ref{Oper2}), the operators can be written as\\
\bea
\label{OS2}
{\cal O}_S(y) &=& \sum_D \bar\chi(y)_{D} \, \chi(y)_D\, ,\\
{\cal O}_V(y) &=& \sum_D \bar\chi(y)_{D+\hat{\mu}}\,U_{D+\hat{\mu},D}\,\chi(y)_D\,\eta_\mu(D)\,,\\
{\cal O}_T(y) &=& \frac{1}{i} \sum_D\,\bar\chi(y)_{D+\hat{\mu}+\hat{\nu}}\,U_{D+\hat{\mu}+\hat{\nu},D} \,\chi(y)_D \,\eta_\nu(D) \,\eta_{\mu}(D+\hat{\nu})\,,\label{OT2}\\
{\cal O}_A(y) &=& \sum_D\bar\chi(y)_{D+\hat{\mu}+(1,1,1,1)}\,U_{D+\hat{\mu}+(1,1,1,1),D}\,\chi(y)_D\,\eta_\mu(D)\,\eta_1(D+\hat{\mu})\,\eta_2(D+\hat{\mu})\,\eta_3(D+\hat{\mu})\,\eta_4(D+\hat{\mu})\,,\\
{\cal O}_P(y) &=& \sum_D \bar\chi(y)_{D+(1,1,1,1)}\,U_{D+(1,1,1,1),D}\,\chi(y)_D\,\eta_1(D)\,\eta_2(D)\,\eta_3(D)\,\eta_4(D)\,.
\eea
With the exception of the scalar operator, the remaining operators
contain averages of products of up to four links (in orthogonal
directions) between the fermion and the antifermion fields. For
example, the average entering the tensor operator of Eq.~(\ref{OT2})
is
\be
U_{D+\hat{\mu}+\hat{\nu}, D} = \frac{1}{2}\left[\,
\tilde{\tilde{U}}^\dagger_\nu(ay+aD+a\hat{\mu})\,\,\tilde{\tilde{U}}_\mu^\dagger(ay+aD)
+ \{ \mu\leftrightarrow \nu\}
\,\right]\,,
\label{paths}
\ee
valid when $(D+\hat{\mu}+\hat{\nu})_i \ge D_i$, $i=1,2,3,4$, and
similarly for all other cases.

\section{Calculation of Green's functions}

In this section we describe some of the technical aspects of the
calculation and present our results for one-loop Green's functions. As
a starting point one must derive the vertices for the staggered action
and the operators, up to two gluons, as required in our one-loop
computation. For this reason one may use an equivalent expression of
$\eta_\mu(x)$ appearing in the action  
\be
\eta_\mu(x)=e^{i\pi\bar\mu\,n}\,,\quad x=an\,,\quad \bar\mu=\sum_{\nu=1}^{\mu-1} \hat{\nu}\,.
\ee
Using this form of $\eta_\mu(x)$, instead of the definition of
Eq.~(\ref{equalities}), simplifies the expression for ${\cal
  O}_{\Gamma}$ in terms of Fourier transformed fields,
$\tilde{\chi}(k),\,\tilde{A}_\rho (k)\equiv \tilde{A}^c_\rho(k)\,T^c$: 

 \begin{eqnarray}
{\cal O}_{\Gamma} \hspace{-0.1cm}&=&\hspace{-0.1cm} \int^{\pi}_{-\pi} \frac{d^4 k_1}{(2\pi)^4}\,
\int^{\pi}_{-\pi} \frac{d^4 k_2}{(2\pi)^4} \, \tilde{\bar{\chi}}
(k_1) \, V_{\Gamma}(k_1,k_2) \,\tilde{\chi} (k_2)\nonumber \\[2ex]
&+&\sum_{c,\rho}\,\int^{\pi}_{-\pi} \frac{d^4 k_1}{(2\pi)^4}\, \int^{\pi}_{-\pi}
\frac{d^4 k_2}{(2\pi)^4} \, \int^{\pi}_{-\pi} \frac{d^4 k_3}{(2\pi)^4}
\,
\tilde{\bar{\chi}}(k_1)\,V^{c,\rho}_{\Gamma}(k_1,k_2,k_3;\omega_1,\omega_2)
\,\tilde{\chi}(k_2)\,\tilde{A}^c_\rho (k_3)\nonumber \\[3ex]
&+& \,{\rm 2-gluon\,\,terms} + \cdots
\end{eqnarray}

Thus, after Fourier transformation, the quark-antiquark vertices of Eqs.~(\ref{OS2}) - (\ref{OT2}),
without stout smearing, become
\begin{eqnarray}
V_S(k_1,k_2) &=& \delta\left(k_2 - k_1 \right)\\[3ex]
V_V(k_1,k_2) &=& \delta\left(k_2 - k_1 + \pi \bar{\mu}\right) e^{-i
  k_1\mu}\,,\\[3ex]
V_T(k_1,k_2) &=& \delta\left( k_2 - k_1 + \pi \bar{\mu} + \pi
\bar{\nu}\right) e^{-i k_1\mu}e^{-i k_1\nu}\,\,(\nu>\mu)\,,\\[1ex]
V_A(k_1,k_2) &=& \eta_\mu(\bar{\mu} )\delta\left( k_2 - k_1 + \pi \sum_{\nu=1}^{4}\bar{\nu} + \pi
\bar{\mu} \right) e^{-i (k_{1_1}+ k_{1_2} +k_{1_3} + k_{1_4} - k_{1_\mu})}\,,\\
V_P(k_1,k_2) &=& \delta \left( k_2 - k_1  + \pi \sum_{\nu=1}^{4}\bar{\nu}\right) 
e^{-i (k_{1_1}+ k_{1_2} +k_{1_3} + k_{1_4})}\,.
\end{eqnarray}

As for vertices containing gluons, we give here as an example the
one-gluon vertex of the vector operator, including double stout smearing:
\begin{eqnarray}
V_V^{c,\rho}(k_1,k_2,k_3;\omega_1,\omega_2)\hspace{-0.1cm}
&=&\hspace{-0.1cm} i g T^c  \Bigg[ \cos\left(\frac{k_{3\mu}}{2}+k_{1\mu} \right)
  \delta\left( k_3 - k_2 + k_1 + \pi \bar{\mu} + \pi \mu \right)+i
  \delta\left( k_3 - k_2 + k_1 + \pi \bar{\mu} \right)
  \sin\left(\frac{k_{3\mu}}{2}+k_{1\mu} \right) \Bigg] \nonumber\\
&&\cdot\Bigg\{4 \sin\left(\frac{k_{3\rho}}{2}\right)
\sin\left(\frac{k_{3\mu}}{2}\right)
 \left(\omega_1 + \omega_2 + 2\,\omega_1\,\omega_2\left(-4 +
 \sum_\sigma \cos\left(k_{1\sigma}\right)\right)\right) \nonumber\\
&+& \delta_{\rho\mu}\left((8\omega_1-1)(8\omega_2-1) + 2
 \sum_\sigma \cos\left(k_{3\sigma}\right)\left(\omega_1 + \omega_2 +
2\,\omega_1\,\omega_2 \left(-8 + \sum_\tau \cos\left(k_{3\tau}\right)\right)\right)\right)\Bigg\}\,,
\end{eqnarray}
where $\mu$ is the index of the inserted Dirac matrix ($\gamma_\mu$) and $\rho$ is the index of
the gluon.

Given that the argument $y$ of the operators ${\cal O}_\Gamma$ runs only over
even integers, summation over the position of ${\cal O}_\Gamma$,
followed by Fourier transformation leads to expressions of the form
\be
\sum_{y_\mu\,\epsilon\,2{\mathbb Z}} e^{i\,y\cdot k} = \frac{1}{16}(2\pi)^4\sum_C
\delta_{2\pi}\left(k + \pi\,C\right)\,,
\ee
where $\delta_{2\pi}\left(k \right)$ stands for the standard periodic
$\delta$ function with nonvanishing support at $k\,{\rm mod}2\pi=0$.
Since contributions to the continuum limit come from the neighborhood
of each of the 16 poles of the external momenta $q$, at $q_\mu =
(\pi/a) C_\mu$, it is useful to define $q'_\mu$ and $C_\mu$ through
\be
q_\mu = q'_\mu+\frac{\pi}{a}C_\mu \quad (\,{\rm mod}(\frac{2\pi}{a})\,),\quad (C_\mu\,\,\epsilon\,\,\{0,1\})\,,
\label{C1}
\ee
where the ``small" (physical) part $q'$ has each of its components
restricted to one-half of the Brillouin zone: $-\pi/(2a) \le q'_\mu
\le \pi/(2a)$. Thus, conservation of external momenta takes the form 
\be
\delta_{2\pi}(a\,q_1 -a\,q_2 +\pi D) =
\frac{1}{a}\delta(q_1' -q_2')\, \prod_\mu \delta_{C_{1\mu}-C_{2\mu}+D_\mu,0}\,.
\ee

For the algebraic operations involved in evaluating the Feynman
diagrams relevant to this calculation, we make use of our symbolic
package in Mathematica; a description of this can be found, e.g.,
in a previous publication~\cite{Constantinou:2009tr}. 

%%%%%%%%%%%%%%%%%%%%%%%%%%%%%%%%%%%%%%%%%%%%%%%%%%%%%%%%%%%%%%%%%%%%%%%%%%%%%%%%%%%%%%%%

\subsection{Fermion propagator}

We compute the one-loop correction to the fermion propagator in order to
obtain the renormalization function of the fermion field, an essential
ingredient for the renormalization of the operators ${\cal O}_\Gamma$.
The tree-level fermion propagator in the basis of the $\chi$
fields can be written as
\be
{\displaystyle{
S_{tree}(q_1,q_2) = (2 \pi)^4\,\, \frac{-\displaystyle\frac{i}{a}\,
\displaystyle\sum_\mu\sin(a\,q_{1\mu}) \delta(q_1- q_2+\displaystyle\frac{\pi\bar\mu}{a})
+ m\,\delta(q_1 - q_2)}{\displaystyle\frac{1}{a^2}\displaystyle\sum_\mu\sin^2(a\,q_{1\mu})+ m^2}\,.
}}
\label{prop}
\ee

\begin{center}
\begin{figure}[!h]
\centerline{\psfig{figure=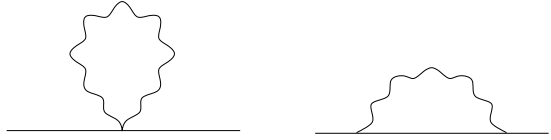,height=1.8truecm}}
\begin{minipage}{10cm}
\caption{One-loop diagrams contributing to the 
fermion propagator. Wavy (solid) lines represent gluons (fermions).}
\label{figprop1}
\end{minipage}
\end{figure}
\end{center}

The one-loop Feynman diagrams that enter the calculation of the two-point,
one-particle irreducible (1PI), amputated Green's function $S^{-1}(p)$
are illustrated in Fig.~\ref{figprop1}. 

\bigskip
We have computed $S^{-1}(p)$ for general values of the gauge
parameter $\alpha$ ($\alpha=0$: Landau gauge, $\alpha=1$: Feynman
gauge), the stout smearing parameters
$\omega_{A_1},\,\omega_{A_2}$, the Lagrangian mass $m$, the number of colors
$N_c$, and the external momenta $q_1$, $q_2$. We have obtained results
using different sets of values for the Symanzik coefficients (shown in
Ref.~\cite{Constantinou:2009tr}). In presenting our result,
Eq.~(\ref{Sinverse}), for $S^{-1}(p)$ up to one loop, the quantities
$e_1,\,e_2$ are numerical coefficients that depend on the Symanzik
coefficients and the stout smearing parameters.
In Appendix~\ref{appB} we provide the general form of $e_1,\,e_2$ and
tabulate their numerical values for the Wilson and tree-level Symanzik
cases; for other actions see Appendix~\ref{appC}. In all expressions
the systematic errors (coming from an extrapolation to infinite lattice
size of our numerical loop integrals) are smaller than the last digit
we present. 
\bea
S^{-1}_{1-loop} &=& \left(\sum_\rho\,\delta(q_1 - q_2 +
\frac{\pi}{a}\bar{\rho})\,i\,p_\rho\,(-1)^{C_{1\rho}}\right)\times\nonumber \\
&&\qquad\Bigg[
        1 + \frac{g^2\,C_F}{16 \pi^2}\,\Big[( e_1 +4.79201\,\alpha 
- \alpha\,\left( \log\left(a^2 m^2 + a^2 p^2\right)+ \frac{m^2}{p^2} -
        \frac{m^4}{p^4}\,\log\left(1+\frac{p^2}{m^2}\right)\right)
        \Big]\Bigg] \nonumber \\
&+&\delta(q_1 - q_2)\, m \times \nonumber\\
&&\qquad\Bigg[ 1 +
        \frac{g^2\,C_F}{16 \pi^2}\,\Big[ e_2 +5.79201\,\alpha
- (3+\alpha)\left( \log\left(a^2 m^2+
        a^2 p^2\right)+ \frac{m^2}{p^2}\,\log\left(1+\frac{p^2}{m^2}\right)\right)\Big]
        \Bigg]\nonumber\\
&+&{\cal O}(a^1)\,,
\label{Sinverse}
\eea
$
\displaystyle
q_1,q_2: {\rm external\,momenta},\quad
C_F \equiv \frac{N_c^2-1}{2N_c},\quad
a\,p_\rho\equiv\left(a\,q_{1\rho}+\frac{\pi}{2}\right)_{{\rm mod}\pi} - \frac{\pi}{2} =
\left(a\,q_{2\rho}+\frac{\pi}{2}\right)_{{\rm mod}\pi} - \frac{\pi}{2}\,,
$
and $C_1$ is defined in Eq.~(\ref{C1}).
Equation ~(\ref{Sinverse}) does have the expected structure of an inverse
propagator, once one identifies, in the continuum limit:
\be
\sum_\rho\,\delta(q_1 - q_2 + \frac{\pi}{a}\bar{\rho})\,p_\rho\,(-1)^{C_{1\rho}}
\,\,\tendto_{a\rightarrow 0}\,\,\delta(q'_1 - q'_2) \qslash_1\,'
\ee
We denote the expression in square brackets in the last line of Eq.~(\ref{Sinverse}) as
$\Sigma_m(q^2,m)$; from this we will extract the multiplicative
renormalization of the Lagrangian mass, $\Zm$.

\subsection{Fermion bilinears}
\begin{figure}[!h]
\begin{center}
\centerline{\psfig{figure=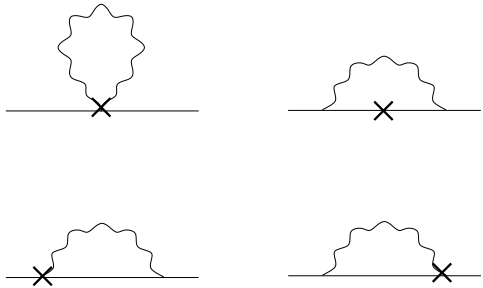,height=3.8truecm}}
%\vspace{-0.3cm}                                                                                                                                                                                          
\begin{minipage}{10cm}
\caption{One-loop diagrams contributing to the fermion-antifermion
  Green's functions of the bilinear operators. A wavy (solid) line
  represents gluons (fermions). A cross denotes an insertion of the
  operator ${\cal O}_\Gamma$.}
\label{figbil2}
\end{minipage}
\end{center}
\end{figure}

In the context of this work we also study the 1PI,
amputated, two-point Green's functions of the operators 
${\cal O}_\Gamma$, defined in Eqs.~(\ref{OS2})$-$(\ref{OT2}) up to
one loop: $\Lambda_{{\cal O}_\Gamma}^{1-loop}$.
The 1PI Feynman diagrams that enter the calculation of the above
Green's functions are shown in Fig.~\ref{figbil2}, and include up to
two-gluon vertices extracted from the operator (the cross in the
diagrams). The appearance of gluon lines on the operator stems from
the product $U_{C,D}$ in the operator definition
[Eq.~(\ref{Oper})]\footnote{For ${\cal O}_S$ only the top right
  diagram of Fig.~\ref{figbil2} contributes, since $U_{C,D}=\openone$.}.

Analogous expressions to Eq.~(\ref{Sinverse}) arise for the bilinears
as well. We note that the extraction of $Z_{{\cal O}_\Gamma}$ in 
a mass-independent scheme, such as RI$'$-MOM, necessitates evaluation
of $\Lambda_{{\cal O}_\Gamma}^{1-loop}$ for $m=0$ only. Nevertheless,
we have included a nonzero Lagrangian mass to our computations; this
allows us to derive the renormalized Green's functions at $m \neq
0$. Comparing the latter with results using a different regularization
scheme (e.g. dimensional regularization) provides another check in our
computation.
 
Although computing the diagrams of Fig.~\ref{figbil2} does not use
the expression of the propagator [Eq.~(\ref{Sinverse})], all our
results shown in Eqs.~(\ref{LambdaS}) $-$ (\ref{LambdaP}) are expressed
in terms of $e_1$ [see Eqs.~(\ref{Sinverse}) and (\ref{e1})]. The reason
for that is to show explicitly the contribution of the quantities
$\lambda_{\cal O}$ [Eqs.~(\ref{ZS}) - (\ref{ZP})] which appear in the
renormalization functions $Z_{\cal O}$ [Eqs.~(\ref{ZcalS}) - (\ref{ZcalP})].

Dropping an overall Dirac $\delta$ function of momentum conservation,
and denoting the physical momentum of the fermion and antifermion by
$p$, we obtain $\Lambda^{1-loop}_{\cal O}$
\bea
\Lambda^{1-loop}_S = \openone + \frac{g^2\,C_F}{16 \pi^2}
\hspace{-0.1cm}&\Bigg[&\hspace{-0.1cm}
e_1 - \lambda_S + 5.79201 \,\alpha +
i\,\pslash\, \left(4\alpha\,\frac{m^3}{(p^2)^2} \log
\left(1+\frac{p^2}{m^2}\right)-4\alpha\,\frac{m}{p^2}\right)\nonumber\\
&-&(\alpha+3)\left(3\,\frac{m^2}{p^2}\log\left(1+\frac{p^2}{m^2}\right)
+ \log \left(a^2 m^2+a^2 p^2\right)   \right)
\Bigg]
\label{LambdaS} \\ [3ex]
\Lambda^{1-loop}_V = \gamma_\mu+ \frac{g^2\,C_F}{16 \pi^2}
\hspace{-0.1cm} &\Bigg[& \hspace{-0.1cm}
\gamma_\mu\left(e_1 - \lambda_V + 4.79201\,\alpha
  -\alpha
  \frac{m^2}{p^2}
-\alpha  \log \left(a^2 m^2+a^2 p^2\right) +\alpha
\frac{m^4}{(p^2)^2}\log\left(1+\frac{p^2}{m^2}\right)\right)\nonumber\\
&+&i p_\mu \left(2 \alpha \frac{m}{p^2}+6 
   \frac{m}{p^2}-\left(2 \alpha \frac{m^3}{(p^2)^2}+6 
   \frac{m^3}{(p^2)^2}\right)\log
   \left(1+\frac{p^2}{m^2}\right)\right)\nonumber\\
&-&\pslash p_\mu \left(2 \alpha
    \frac{1}{p^2} - 4 \alpha\frac{m^2}{(p^2)^2}+ 4
   \alpha  \frac{m^4}{(p^2)^3} \log
   \left(1+\frac{p^2}{m^2}\right)\right) \Bigg]
\label{LambdaV} \\[3ex]
\Lambda^{1-loop}_T = \sigma_{\mu\nu} + \frac{g^2\,C_F}{16 \pi^2}
\hspace{-0.1cm}&\Bigg[&\hspace{-0.1cm}
\gamma_\mu\, \gamma_\nu\,\Bigg(e_1 - \lambda_T  + 3.79201\,\alpha - (1-\alpha)
\Big(2\,\frac{m^2}{p^2} - \log \left(a^2 m^2+a^2 p^2\right)\nonumber\\
&&\phantom{\gamma_\mu\, \gamma_\nu\,\Bigg(}-\left(2\,\frac{m^4}{(p^2)^2}+\frac{m^2}{p^2}\right) \log
\left(1+\frac{p^2}{m^2}\right)
\Big)\Bigg)\nonumber\\
&-&(\gamma_\mu \pslash p_\nu - \gamma_\nu\pslash p_\mu)
(1-\alpha)\left(\left(4\,\frac{m^4}{(p^2)^3}+2\,\frac{m^2}{(p^2)^2}\right)\log
\left(1+\frac{p^2}{m^2}\right)-4\,\frac{m^2}{(p^2)^2}\right)\nonumber\\
&-&i\,\gamma_\mu\, \gamma_\nu\pslash\left(4\,\frac{m^3} {(p^2)^2}\log
\left(1+\frac{p^2}{m^2}\right)-4\,\frac{m}{p^2}\right)\nonumber\\
&-&i\,(\gamma_\mu p_\nu - \gamma_\nu p_\mu)
\left(4\,\frac{m}{p^2}-4\,\frac{m^3}{(p^2)^2}\log\left(1+\frac{p^2}{m^2}\right)\right)
\Bigg] \label{LambdaT} %\\ [5ex]
\eea

\bea
\Lambda^{1-loop}_A = \gamma_5\,\gamma_\mu + \frac{g^2}{16 \pi^2} C_F  \gamma_5\hspace{-0.1cm}
&\Bigg[&\hspace{-0.1cm} \gamma_\mu \Bigg(e_1- \lambda_A +4.79021\,\alpha
- (2-\alpha)\frac{m^2}{p^2} - \alpha \log \left(a^2 m^2+a^2 p^2\right) \nonumber\\
&&\phantom{ \gamma_\mu \Bigg(}+\left(2\,(1-\alpha)\frac{m^4}{(p^2)^2}- 2\,(1+\alpha)\frac{m^2}{p^2}\right) \log
\left(1+\frac{p^2}{m^2}\right)
\Bigg)\nonumber\\
&-& i\,p_\mu (1-\alpha) \left(2\frac{m}{p^2}-2\frac{m^3}{(p^2)^2}
\log\left(1+\frac{p^2}{m^2}\right)\right)\nonumber\\
&+& i \gamma_\mu \,\pslash\,(1-\alpha)\left(
2\frac{m}{p^2}-2\frac{m^3}{(p^2)^2} \log
\left(1+\frac{p^2}{m^2}\right)\right)\nonumber\\
&-&\pslash p_\mu\Bigg(-8\frac{m^2}{(p^2)^2} +
2 \alpha\frac{1}{p^2}+ 4 \alpha\frac{m^2}{(p^2)^2}\nonumber \\
&&\phantom{\pslash p_\mu\Bigg(}+\left(8\frac{m^4}{(p^2)^3}-4 \alpha
\frac{m^4}{(p^2)^3}+4\frac{m^2}{(p^2)^2}-4 \alpha \frac{m^2}{(p^2)^2}
\right)\log\left(1+\frac{p^2}{m^2}\right)\Bigg)\Bigg]
\label{LambdaA} \\ [3ex]
\Lambda^{1-loop}_P = \gamma_5 + \frac{g^2\,C_F}{16 \pi^2}  \gamma_5
\hspace{-0.1cm}&\Bigg[& \hspace{-0.1cm}e_1 - \lambda_P + 5.79201\,\alpha
-(\alpha +3) \frac{m^2} {p^2}\log \left(1+\frac{p^2}{m^2}\right)
-(\alpha +3) \log \left(a^2 m^2+a^2 p^2\right)
\Bigg]\qquad\qquad \label{LambdaP}
\eea

The quantities $\lambda_{\cal O}$ are independent of the mass, gauge
parameter, external momentum, and lattice spacing; they depend on the
gluon action and the stout parameters. As discussed earlier, we have
employed different parameters for the two smearing steps; in fact, we
have also kept the parameters of the action's smearing procedure
($\omega_{A_{1}},\,\omega_{A_{2}}$) distinct from the parameters of
the operator smearing ($\omega_{{\cal O}_{1}},\,\omega_{{\cal
    O}_{2}}$). For the tree-level Symanzik action and for general
values of the stout parameters we obtained
\bea
\lambda_S &=& -34.3217 + 389.210 \,\left(\omega_{A_1} +
\omega_{A_2}\right)
- 1403.65\,\left(\omega_{A_1}^2 + \omega_{A_2}^2 \right)
- 5614.59\,\omega_{A_1}\,\omega_{A_2}\nonumber\\
&+& 23395.4\,\left(\omega_{A_1}^2\,\omega_{A_2} +
\omega_{A_1}\,\omega_{A_2}^2\right)
- 106814 \,\omega_{A_1}^2\,\omega_{A_2}^2 \,, 
\label{ZcalS} \\[3.5ex]
\lambda_V &=& 86.7568\,\left[\left(\omega_{A_1} +
  \omega_{A_2}\right)\,-\,\left(\omega_{{\cal O}_1} +
  \omega_{{\cal O}_2}\right)\right]
- 337.383\,\left[\left(\omega_{A_1}^2 + \omega_{A_2}^2
  \right)\,-\,\left(\omega_{{\cal O}_1}^2 + \omega_{{\cal O}_2}^2
  \right)\right]\nonumber\label{Zcalv}\\
&-&
1349.53\,\left(\omega_{A_1}\,\omega_{A_2}\,-\,\omega_{{\cal O}_1}\,\omega_{{\cal O}_2}\right)
+ 5950.81\,\left[\left(\omega_{A_1}^2\,\omega_{A_2} +
  \omega_{A_1}\,\omega_{A_2}^2\right)\,-\,\left(\omega_{{\cal O}_1}^2\,\omega_{{\cal O}_2}
  + \omega_{{\cal O}_1}\,\omega_{{\cal O}_2}^2\right)\right]\nonumber\\
&-&
28627.2\,\left(\omega_{A_1}^2\,\omega_{A_2}^2\,-\,\omega_{{\cal O}_1}^2\,\omega_{{\cal O}_2}^2\,,
\right) \\ [3.5ex]
\lambda_T &=& 8.88342 + 116.579\,\left(\omega_{A_1} +
\omega_{A_2}\right)
- 200.588\,\left(\omega_{{\cal O}_1} + \omega_{{\cal O}_2}\right)
- 531.759\,\left(\omega_{A_1}^2 + \omega_{A_2}^2 \right)
\nonumber\\
&+& 780.590\,\left(\omega_{{\cal O}_1}^2 + \omega_{{\cal O}_2}^2 \right)
- 2095.16\,\omega_{A_1}\,\omega_{A_2}
+ 3154.24\,\omega_{{\cal O}_1}\,\omega_{{\cal O}_2}\nonumber\\
&+& 31.8743\,\left(\omega_{A_1} +
\omega_{A_2}\right)\,\left(\omega_{{\cal O}_1} +
\omega_{{\cal O}_2}\right)
+ 9877.233\,\left(\omega_{A_1}^2\,\omega_{A_2} +
\omega_{A_1}\,\omega_{A_2}^2\right)\nonumber\\
&-& 13993.1\,\left(\omega_{{\cal O}_1}^2\,\omega_{{\cal O}_2} +
\omega_{{\cal O}_1}\,\omega_{{\cal O}_2}^2\right)
- 284.001\,\Big(\left(\omega_{A_1} +
  \omega_{A_2}\right)\omega_{{\cal O}_1}\,\omega_{{\cal O}_2}\nonumber\\
&+&\,\omega_{A_1}\,\omega_{A_2}\,\left(\omega_{{\cal O}_1} +
  \omega_{{\cal O}_2}\right)\Big)
- 48519.3\,\omega_{A_1}^2\,\omega_{A_2}^2     \nonumber\\
&+& 68237.1 \,\omega_{{\cal O}_1}^2\,\omega_{{\cal O}_2}^2
+ 2709.49
\,\omega_{A_1}\,\omega_{A_2}\omega_{{\cal O}_1}\,\omega_{{\cal O}_2}\,,\\[3.5ex]
\lambda_A &=& 17.0363 + 117.584\,\left(\omega_{A_1} +
\omega_{A_2}\right)
- 314.355\,\left(\omega_{{\cal O}_1} + \omega_{{\cal O}_2}\right)
- 518.419\,\left(\omega_{A_1}^2 + \omega_{A_2}^2 \right)
\nonumber\\
&+& 1223.79\,\left(\omega_{{\cal O}_1}^2 + \omega_{{\cal O}_2}^2 \right)
- 2041.80\,\omega_{A_1}\,\omega_{A_2}
+ 4927.06\,\omega_{{\cal O}_1}\,\omega_{{\cal O}_2}\nonumber\\
&+& 31.8758\,\left(\omega_{A_1} +
\omega_{A_2}\right)\,\left(\omega_{{\cal O}_1} +
\omega_{{\cal O}_2}\right)
+ 9559.98\,\left(\omega_{A_1}^2\,\omega_{A_2} +
\omega_{A_1}\,\omega_{A_2}^2\right)\nonumber\\
&-& 21823.5\,\left(\omega_{{\cal O}_1}^2\,\omega_{{\cal O}_2} +
\omega_{{\cal O}_1}\,\omega_{{\cal O}_2}^2\right)
- 210.274\,\Big(\left(\omega_{A_1} +
\omega_{A_2}\right)\omega_{{\cal O}_1}\,\omega_{{\cal O}_2}\nonumber\\
&+&\,\omega_{A_1}\,\omega_{A_2}\,\left(\omega_{{\cal O}_1} +
\omega_{{\cal O}_2}\right)\Big)
- 47154.2\,\omega_{A_1}^2\,\omega_{A_2}^2     \nonumber\\
&+& 105754.\, \,\omega_{{\cal O}_1}^2\,\omega_{{\cal O}_2}^2
+ 1396.94
\,\omega_{A_1}\,\omega_{A_2}\omega_{{\cal O}_1}\,\omega_{{\cal O}_2} \,,%\\[5ex]
\eea
\bea
\lambda_P &=& 25.7425 + 119.062\,\left(\omega_{A_1} +
\omega_{A_2}\right)
- 428.120\,\left(\omega_{{\cal O}_1} + \omega_{{\cal O}_2}\right)
- 518.541\,\left(\omega_{A_1}^2 + \omega_{A_2}^2 \right)
\nonumber\\
&+& 1667.00\,\left(\omega_{{\cal O}_1}^2 + \omega_{{\cal O}_2}^2 \right)
- 2042.29\,\omega_{A_1}\,\omega_{A_2}
+ 6699.88\,\omega_{{\cal O}_1}\,\omega_{{\cal O}_2}\nonumber\\
&+& 31.8765\,\left(\omega_{A_1} +
\omega_{A_2}\right)\,\left(\omega_{{\cal O}_1} +
\omega_{{\cal O}_2}\right)
+ 9435.40\,\left(\omega_{A_1}^2\,\omega_{A_2} +
\omega_{A_1}\,\omega_{A_2}^2\right)\nonumber\\
&-& 29654.0\,\left(\omega_{{\cal O}_1}^2\,\omega_{{\cal O}_2} +
\omega_{{\cal O}_1}\,\omega_{{\cal O}_2}^2\right)
- 210.274\,\Big(\left(\omega_{A_1} +
\omega_{A_2}\right)\omega_{{\cal O}_1}\,\omega_{{\cal O}_2}\nonumber\\
&+&\,\omega_{A_1}\,\omega_{A_2}\,\left(\omega_{{\cal O}_1} +
\omega_{{\cal O}_2}\right)\Big)
- 44803.9\,\omega_{A_1}^2\,\omega_{A_2}^2     \nonumber\\
&+& 143482.\, \,\omega_{{\cal O}_1}^2\,\omega_{{\cal O}_2}^2
+ 1657.76
\,\omega_{A_1}\,\omega_{A_2}\omega_{{\cal O}_1}\,\omega_{{\cal O}_2}\,.
\label{ZcalP}
\eea
We note in passing that in the absence of stout smearing
($\omega_{A_i}=\omega_{{\cal O}_i}=0$) $\lambda_V=0$, which implies
that $\ZV^{\rm RI'}=\ZV^{\rm \overline{MS}}=1$ [cf. Eqs.~(\ref{ZV}),
(\ref{MS_conversion_VA})], as is well known from current
conservation. In addition, Eq.~(\ref{Zcalv}) shows that
nonrenormalization of ${\cal O}_V$ applies also when
$\omega_{A_i}=\omega_{{\cal O}_i}$; this follows from the fact that
the stout link version of ${\cal O}_V$ mimics that of the action, and
thus current conservation applies equally well in this case.

The dependence of the Green's functions of Eqs.~(\ref{LambdaS}) $-$
(\ref{LambdaP}) on mass and external momentum is regularization
independent and agrees for instance with the results of 
Refs.~\cite{Capitani:2000xi,Alexandrou:2012mt}.
As is well known, in the limit of zero mass the vector and axial
Green's functions beyond tree level are not multiples of their
tree-level values: There appear additional, finite contributions with
tensor structures which are distinct from those at tree level. These
contributions denoted as $\Sigma^{(2)}_V$ and $\Sigma^{(2)}_A$ can
be read off Eqs.~(\ref{LambdaV}) and ~(\ref{LambdaA}):
\bea
\Sigma^{(2)}_V &=&  \frac{g^2\,C_F}{16 \pi^2} \Big[-2\,\alpha\frac{\qslash\,q_\mu}{q^2} \Big] \\
\Sigma^{(2)}_A &=&  \frac{g^2\,C_F}{16 \pi^2} \Big[-2\,\alpha\frac{\gamma_5\,\qslash\,q_\mu}{q^2} \Big] 
\eea
A similar contribution for the tensor bilinear does not appear up to,
and including, three loops~\cite{Gracey:2003yr}.
The role of $\Sigma^{(2)}_V$ and $\Sigma^{(2)}_A$ in the renormalization of
${\cal O}_V$ and ${\cal O}_A$ will be discussed in the next section.

\section{Renormalization functions}

\subsection{Fermion field and fermion bilinear renormalization functions in the RI$'$-MOM scheme}

RFs for operators and action parameters
relate bare quantities regularized on the lattice, to their
renormalized continuum counterparts:
\be
\psi^{\rm renorm} = \Zq^{\frac{1}{2}} \psi^{\rm bare}\,,\qquad
m^{\rm renorm} = \Zm m^{\rm bare}\,,\qquad
{\cal O}_\Gamma^{\rm renorm} = Z_{{\cal O}_\Gamma} {\cal O}_\Gamma^{\rm bare}\,.
\ee
The RFs of lattice operators are necessary ingredients in the
prediction of physical probability amplitudes from lattice matrix
elements. In this section we present the multiplicative RFs in the
RI$'$-MOM scheme of the fermion field ($\Zq$), the fermion mass ($\Zm$),
and the fermion bilinears.

The RI$'$-MOM renormalization scheme consists in requiring that the
forward amputated Green's function $\Lambda(p)$ computed in the chiral
limit and at a given (large Euclidean) scale $p^2=\mu^2$ be equal to
its tree-level value. Our results for the RFs are presented for
arbitrary values of the renormalization scale $\mu$.
This requirement leads to the following definitions for
$\Zq^{{\rm RI}'},\,\Zm^{{\rm RI}'},\,Z_{{\cal O}_\Gamma}^{{\rm RI}'}$:
\begin{eqnarray}
S_{1-loop}^{-1}\,\Big{|}_{p^2=\mu^2,\,\, m=0}  &=& S_{tree}^{-1}\,\Big{|}_{p^2=\mu^2,\,\, m=0}\, \Zq^{{\rm RI}'}(\mu)\,,\\[2ex]
 \Sigma_m\,\Big{|}_{p^2=\mu^2,\,\, m=0} &=&\Zm^{{\rm RI}'}(\mu) \, \Zq^{{\rm RI}'}(\mu)\,,\\[2ex]
\Lambda_{{\cal O}_\Gamma}^{1-loop}\,\Big{|}_{p^2=\mu^2,\,\, m=0} &=& \Lambda_{{\cal O}_\Gamma}^{tree}
\,\Zq^{{\rm RI}'}(\mu)\,\left(Z^{{\rm RI}'}_{{\cal
    O}_\Gamma}(\mu)\right)^{-1}\,, \quad (\Gamma=S,\,T,\,P)\,,
\label{renormalization cond}
\end{eqnarray}
where $S_{tree}^{-1}$ is the tree-level result for the inverse
propagator, and $\Lambda_{{\cal O}_\Gamma}^{tree}$ is the tree-level
value of the Green's function for ${\cal O}_\Gamma$. 

The presence of $\Sigma^{(2)}_V$ and $\Sigma^{(2)}_A$ in the one-loop
Green's functions of ${\cal O}_V$ and ${\cal O}_A$ makes a
prescription such as Eq.~(\ref{renormalization cond}) inapplicable in
those cases. Instead we employ
\be
\left(\Lambda_{V,\,A}^{1-loop} - \Sigma^{(2)}_{V,\,A}\right)\,\Bigg{|}_{p^2=\mu^2,\,\, m=0} = 
\Lambda_{V,\,A}^{tree}
\Zq^{{\rm RI}'}(\mu)\,\left(Z^{{\rm RI}'}_{V,\,A}(\mu)\right)^{-1}\,,
\label{VArenormCondition}
\ee
and thus take into account only the terms in $\Lambda_{V,\,A}$ which
are proportional to their corresponding tree-level values.

The expressions we obtain using our results for
$\Lambda_{{\cal O}_\Gamma}^{1-loop}$ are shown here only for the
tree-level improved Symanzik gauge action. The quantities
$\lambda_{\cal O}$ are defined in Eqs.~(\ref{ZcalS}) $-$
(\ref{ZcalP}). We note that the results for $\Zm$ and $\ZS$ are related
by $\Zm = \ZS^{-1}$ as expected,

%%%%%%%%% Zq %%%%%%%%%                                                                                                                                                                                    
\vspace*{0.5cm}
\begin{eqnarray}
 \Zq^{\rm RI'} = 1 + \frac{g^2\,C_F}{16 \pi^2}\,\Bigl[\hspace{-0.1cm}
&-&\hspace{-0.1cm} \alpha \,\log\left(a^2\,\mu^2\right) +
   4.79201\,\alpha\, - 7.21363 +
 124.515\,\left(\omega_{A_1} + \omega_{A_2}\right)
- 518.433\,\left(\omega_{A_1}^2 + \omega_{A_2}^2 \right) \nonumber\\
&-& 2073.73\,\omega_{A_1}\,\omega_{A_2} + 
9435.35\,\left(\omega_{A_1}^2\,\omega_{A_2}
+ \omega_{A_1}\,\omega_{A_2}^2\right)
- 45903.1\,\omega_{A_1}^2\,\omega_{A_2}^2 \Bigr]\,,
\end{eqnarray}

\begin{eqnarray}
\Zm^{\rm RI'} = 1 + \frac{g^2\,C_F}{16 \pi^2}\,
\Bigl[\hspace{-0.1cm}&-&\hspace{-0.1cm}
3\,\log\left(a^2\,\mu^2\right) +\,\alpha\, +
34.3217  - 389.210 \,\left(\omega_{A_1} +
\omega_{A_2}\right)
+ 1403.65\,\left(\omega_{A_1}^2 + \omega_{A_2}^2 \right) \nonumber\\
&+& 5614.59\,\omega_{A_1}\,\omega_{A_2} -
23395.4\,\left(\omega_{A_1}^2\,\omega_{A_2} +
\omega_{A_1}\,\omega_{A_2}^2\right)
+ 106813.\, \,\omega_{A_1}^2\,\omega_{A_2}^2 \Bigr]\,,
\end{eqnarray}

\bea
\ZS^{\rm RI'} &=& 1 + \frac{g^2\,C_F}{16 \pi^2}\,
\Bigl[\lambda_S -\,\alpha\, + 3\,\log\left(a^2\,\mu^2\right) \Bigr]\,,
\label{ZS}\\[4ex]
\ZV^{\rm RI'}&=& 1 + \frac{g^2\,C_F}{16\pi^2}\,\Bigl[\lambda_V \Bigr]\,,\\[4ex]
\label{ZV}
\ZT^{\rm RI'} &=& 1 + \frac{g^2\,C_F}{16\pi^2}\,\Bigl[\lambda_T  + \,\alpha
-\,\log\left(a^2\,\mu^2\right) \Bigr]\,,\\[4ex]
\ZA^{\rm RI'} &=& 1 + \frac{g^2\,C_F}{16 \pi^2}\,\Bigl[\lambda_A \Bigr]\,,\\[4ex]
\ZP^{\rm RI'} &=& 1 + \frac{g^2\,C_F}{16 \pi^2}\,\Bigl[\lambda_P - \,\alpha
+3\,\log\left(a^2\,\mu^2\right) \Bigr]\,.
\label{ZP}
\eea

In order to compare perturbative and nonperturbative estimates of RFs
one clearly needs to employ the same renormalization prescription in
both cases. In the context of a numerical simulation the term
$\Sigma^{(2)}$ for the vector and axial cases is often not removed
from the Green's functions, in contrast to what is done perturbatively
in Eq.~(\ref{VArenormCondition}). Therefore, an alternative RI$'$
renormalization prescription appears more natural:
\be
Z_q^{-1}\,Z_{V,A}^{\rm RI'\,alter}\,\,
{\rm Tr}\Big[\Lambda_{V,A}^{1-loop}\,\,\Lambda_{V,A}^{tree} \Big] = 
{\rm Tr}\Big[\Lambda_{V,A}^{tree}\,\,\Lambda_{V,A}^{tree} \Big]\,.
\label{RIalter}
\ee
Using the above prescription, the extracted $\ZV^{\rm RI'\,alter}$ and
$\ZA^{\rm RI'\,alter}$ take the form
\bea
\ZV^{\rm RI'\,alter} &=& \ZV^{\rm RI'}  + \frac{g^2\,C_F}{16 \pi^2}\,\frac{\alpha}{2}\,, \\[4ex]
\ZA^{\rm RI'\,alter} &=& \ZA^{\rm RI'}  + \frac{g^2\,C_F}{16 \pi^2}\,\frac{\alpha}{2}\,.
\eea

\subsection{Conversion to the $\overline{\rm MS}$ scheme}

In this section we provide the expressions for the RFs in the
$\overline{\rm MS}$ continuum scheme using conversion factors
adapted from Ref.~\cite{Gracey:2003yr}. These conversion factors do
not depend on the regularization scheme (and, thus, they are
independent of the lattice discretization) when expressed in terms of
the renormalized coupling constant. However, expressing them in terms
of the bare coupling constant introduces a dependence on the action.
In our analysis we use one-loop formulas, which are action
independent. The definition for the conversion factors $C_{\cal O}$
is as follows:
\be
Z_{{\cal O}_\Gamma}^{\overline{\rm MS},\,\rm NDR} = C_{\cal O}\,Z_{{\cal O}_\Gamma}^{\rm RI'}\,.
\label{Conv}
\ee
The above conversion factors refer to the naive dimensional
regularization (NDR) of the $\overline{\rm MS}$ scheme (see e.g.,
Ref.~\cite{Buras:1989xd}), in which $C_P=C_S$ and $C_A=C_V$. From
Eq.~(\ref{Conv}) one obtains \footnote{Note that, at variance with
Eq.~(\ref{MS_conversion_VA}), the conversion factors $C_{V,A}$ will not
be equal to 1 if one uses, e.g., the "alternative" RI$'$
renormalization scheme of Eq.~(\ref{RIalter}).}
\bea
\Zq^{\overline{\rm MS},\,\rm NDR} &=& \Zq^{\rm RI'} -
\frac{g^2\,C_F}{16\pi^2}\,\alpha + {\cal O}(g^4) \,,\\
\label{MS_conversion_SP}
Z_{\rm S,P}^{\overline{\rm MS},\,\rm NDR} &=& Z_{\rm S,P}^{\rm RI'} +
\frac{g^2\,C_F}{16\pi^2}\,(4+\alpha) + {\cal O}(g^4)\,,\\
\label{MS_conversion_VA}
Z_{\rm V,A}^{\overline{\rm MS},\,\rm NDR} &=& Z_{\rm V,A}^{\rm RI'} \,,\\
\ZT^{\overline{\rm MS},\,\rm NDR} &=& \ZT^{\rm RI'} -
\frac{g^2\,C_F}{16\pi^2}\,\alpha + {\cal O}(g^4)\,.
\eea
Other modified minimal subtraction schemes are
related to NDR via additional finite renormalization and affect the
operators which include a $\gamma_5$, due to the nonunique
generalization of $\gamma_5$ to D dimensions. Thus, the treatment of
the pseudoscalar and axial operators in the $\overline{\rm MS}$ scheme
requires special attention. The $\overline{\rm MS}$ renormalized
pseudoscalar and axial operators, as defined in the scheme of 't Hooft
and Veltman (HV)~\cite{'tHooft:1972fi}, involve extra finite factors, $Z_5^P,\,Z_5^A$, in
addition to the conversion factors of
Eqs.~(\ref{MS_conversion_SP}) and (\ref{MS_conversion_VA})~\cite{Larin:1993tq}: 
\bea
Z_5^P &=& 1-\frac{g^2}{16\pi^2}\,(8\,C_F)\,,\\
Z_5^A &=& 1-\frac{g^2}{16\pi^2}\,(4\,C_F)\,.
\eea
The relation between the NDR and the HV schemes is
\bea
\ZP^{\overline{\rm MS},\,\rm HV} &=& \ZP^{\overline{\rm MS},\,\rm NDR}\,Z_5^P\,,\\
\ZA^{\overline{\rm MS},\,\rm HV} &=& \ZA^{\overline{\rm MS},\,\rm NDR}\,Z_5^A\,.
\eea
We would like to point out that although the expressions for $Z_5^A$ and
$Z_5^P$ are, in general, different for flavor singlet and nonsinglet operators,
at one-loop level they coincide.

Other variants of $\overline{\rm MS}$ include the $\rm
DR\overline{EZ}$ and DRED schemes; the conversion from one scheme to
another can be found in Sec. 4 of Ref.~\cite{Patel:1992vu}.
Our results for the fermion bilinears using the Wilson gauge action
and without stout smearing  converted in the $\rm DR\overline{EZ}$
scheme agree with the corresponding results of Ref.~\cite{Patel:1992vu}.

Having obtained $Z_{{\cal O}_\Gamma}^X$ in some renormalization scheme
[$X=({\rm RI}'),\,({\rm RI}'\rm {alter}),\,(\overline{\rm MS},{\rm NDR}),\,
(\overline{\rm MS},{\rm HV})$, etc.] the expression for the renormalized
Green's functions in that scheme $\Lambda_{{\cal O}_\Gamma}^{\rm renorm,X}(p,m)$ 
follow immediately:
\be
\Lambda_{{\cal O}_\Gamma}^{{\rm renorm},X}(p,m) = 
\Lambda_{{\cal O}_\Gamma}^{{\rm bare},X}(p,m)\,\left(Z_q^X\right)^{-1}\,Z_{{\cal O}_\Gamma}^X\,.
\ee

\section{Summary}

In this paper we presented the calculation of the fermion propagator
and the Green's functions for the ultralocal fermion bilinear operators:
scalar, pseudoscalar, vector, axial and tensor. The computations were
performed to one loop in lattice perturbation theory, using
staggered fermions and Symanzik improved gluons parameterized by three
independent ``Symanzik'' coefficients; explicit results are presented
for some of the most commonly used actions in this family: Wilson,
tree-level Symanzik, tadpole improved L\"uscher-Weisz, Iwasaki, and DBW2.

The novelty in our calculations was the stout smearing of the links
that we applied in both the fermion action and in the bilinear
operators. More precisely, we use two steps of stout smearing with
distinguishable parameters. To make our results as general as possible
we also distinguished between the stout parameters appearing in the fermion
action and in the bilinears.

Our expressions for the fermion propagator and the Green's functions
of the bilinear operators exhibit a rather nontrivial dependence on
the external momentum ($q$) and the fermion mass ($m$), and they are
polynomial functions of the gauge parameter ($\alpha$), stout parameters
($\omega_{A_i},\,\omega_{{\cal O}_i}$), and coupling constant
($g$). The numerical coefficients appearing in these expressions
depend on the Symanzik parameters of the gluon action and were presented
for the tree-level Symanzik improved gluon action; the most general
expressions can be found in the electronic document in the form of a
Mathematica input file, Staggered.m\,.

Using the aforementioned results we extract the renormalization
function of the fermion field and those of the fermion bilinears in
the RI$'$-MOM scheme and we provide the appropriate conversion factors
to the $\overline{\rm MS}$ scheme; we pay particular attention to the
operators which include a $\gamma_5$ in their definition. Moreover, for
the case of the vector and axial operators we give an alternative
prescription to obtain the renormalizations in the RI$'$ scheme.

There are several directions in which the present work could be
extended
\begin{itemize}
\item A natural extension would be the computation of the Green's
  functions for operators including covariant derivatives, such as the
  one-derivative vector and axial operators\footnote{Curly brackets
  denote symmetrization and subtraction of the trace.}: 
$\overline \psi \gamma_{\{\mu}\overleftrightarrow D_{\nu\}} \psi$, 
$\,\overline \psi \gamma_5\gamma_{\{\mu}\overleftrightarrow D_{\nu\}} \psi$.
The corresponding renormalization functions may be applied to the
nonperturbative lattice evaluation of the momentum fraction of the
nucleon, $<x>_q$, and the moment of the polarized quark distribution
of the nucleon, $<x>_{\Delta q}$. 
\item A related further work using staggered fermions with stout
  improvement would be a computation of Green's functions for four-Fermi
  operators; a work in this direction can be found in Ref.~\cite{Kim:2011pz}. 
\item A possible improvement to the action may involve further
  iterations of stout smearing; such a procedure has been applied to
  clover~\cite{Capitani:2006ni} and Wilson fermions~\cite{Borsanyi:2011kg}.
\item It would be also interesting to calculate the Green's functions up to
second order in the lattice spacing; such an extension would not only
be useful to constructing improved versions of the operators but
also to remove ${\cal O}(g^2\,a^2)$ contributions from the nonperturbative
estimates of the renormalization functions. Similar computations have
been performed recently with Wilson/clover/twisted mass fermions 
\cite{Constantinou:2009tr, Alexandrou:2010me, Constantinou:2013ada}.
\end{itemize}

\vspace*{1cm}
{\bf{Acknowledgments:}}
This work was partly supported by funding received from the
Cyprus Research Promotion Foundation Contract No.
number TECHNOLOGY/$\Theta$E$\Pi$I$\Sigma$/0311(BE)/16 and  No.
NEA Y$\Pi$O$\Delta$OMH/$\Sigma$TPATH/0308/31. We acknowledge many
helpful discussions with G. Bali, F. Bruckmann, and G. Endr\H{o}di.

\newpage
\appendix
\section{STOUT SMEARING OF THE LINKS}
\label{appA}

Here we present the doubly stout one-gluon link, $U^{(1)}$, for general
values of $\omega_1$ and $\omega_2$, as well as the two-gluon link,
$U^{(2)}$ (only for $\omega_2=0$, to simplify the latter's lengthy
expression):

\begin{eqnarray}
\tilde{\tilde{U}}^{(1)}_{\mu}(x;\omega_1,\omega_2) =
i g \, \Bigl[
&+& A_\mu(x)+ (\omega_1+\omega_2)\,\left(-8A_\mu(x) + \sum_{\rho=\pm 1}^{\pm 4} \Big(A_\mu(x+ a \hat{\rho}) + A_\rho(x) - A_\rho(x + a \hat{\mu})\Big)\right)\nonumber\\
&+& (\omega_1\,\omega_2)\Big[64 A_\mu(x) + \sum_{\rho=\pm 1}^{\pm 4}\Big(-
16 A_\mu(x + a \hat{\rho})  - 8 A_\rho(x) + 8 A_\rho(x + a \hat{\mu}) \Big)\nonumber\\
&+& \sum_{\rho=\pm 1}^{\pm 4}\sum_{\sigma=\pm 1}^{\pm 4}\Big( A_\rho(x +a \hat{\sigma})
-  A_\rho(x + a \hat{\mu} + a \hat{\sigma}) + A_\mu(x + a \hat{\rho} + a \hat{\sigma})\Big)\Big]
\Bigr],
\label{Ut}
\end{eqnarray}

\begin{eqnarray}
\tilde{U}^{(2)}_{\mu}(x;\omega_1,\omega_2=0) =
g^2 \, \Bigl[
&-& \frac{A_\mu(x)^2}{2} + \omega_1
\left(8 A_\mu(x)^2 - \sum_{\rho=\pm 1}^{\pm 4} A_\mu(x) (A_\mu(x + a \hat{\rho})+ A_\rho(x) - A_\rho(x +a \hat{\mu}) )\right)\nonumber\\
&+& \omega_1^2 \Big[-32 A_\mu(x)^2 + \sum_{\rho=\pm 1}^{\pm 4}\Big(8 A_\mu(x) (A_\mu(x+ a \hat{\rho}) + A_\rho(x) - A_\rho(x +a \hat{\mu}) )\Big) \nonumber\\
&+& \sum_{\rho=\pm 1}^{\pm 4}\sum_{\sigma=\pm 1}^{\pm 4}\Big(-\frac{1}{2} A_\mu(x + a \hat{\rho}) A_\mu(x + a \hat{\sigma}) - A_\mu(x +a \hat{\rho}) A_\sigma(x)\nonumber\\
&-& \frac{1}{2}A_\rho(x) A_\sigma(x) + \frac{1}{2}A_\rho(x +a \hat{\mu}) A_\sigma(x)
+ \frac{1}{2}A_\rho(x+a\hat{\mu}-a\hat{\rho}) A_\sigma(x +a \hat{\mu})\nonumber\\
&+& A_\mu(x + a \hat{\rho}) A_\sigma(x +a \hat{\mu})- \frac{1}{2} A_\rho(x -a \hat{\rho}) A_\sigma(x +a \hat{\mu})
  \Big)\Big]
\Bigr]
\label{Utt}
\end{eqnarray}
where we define $A_{-\rho}(y) = - A_\rho(y-a \hat{\rho}), \, \rho>0$\,.

\bigskip
Note: The order in which a product of gluon fields appear in $\tilde{U}^{(2)}_{\mu}$
is irrelevant for the particular diagrams which we compute (since
these two gluons are contracted among themselves); we have used this
fact in order to simplify the expression for $\tilde{U}^{(2)}_{\mu}$.

\newpage
\section{NUMERICAL RESULTS FOR THE PROPAGATOR}
\label{appB}

In this appendix we present the numerical coefficients $e_1$
and $e_2$ appearing in Eq.~(\ref{Sinverse}); these are polynomials in
the two stout smearing parameters of the action
($\omega_{A_{1}},\,\omega_{A_{2}}$):
\bea
e_1 &=& e^{(1,1)} 
+ e^{(1,2)}\,(\omega_{A_{1}}+\omega_{A_{2}})
+ e^{(1,3)}\,(\omega_{A_{1}}^2+\omega_{A_{2}}^2)\nonumber\\[2ex]
&+& e^{(1,4)}\,\omega_{A_{1}}\,\omega_{A_{2}}
+ e^{(1,5)}\,(\omega_{A_{1}}^2 \omega_{A_{2}} +\omega_{A_{1}}\,\omega_{A_{2}}^2)
+ e^{(1,6)} \,\omega_{A_{1}}^2 \omega_{A_{2}}^2\,,
\label{e1}
\eea
\vspace{0.2cm}
\bea
e_2 &=& e^{(2,1)} 
+ e^{(2,2)}\,(\omega_{A_{1}}+\omega_{A_{2}})
+ e^{(2,3)}\,(\omega_{A_{1}}^2+\omega_{A_{2}}^2)\nonumber\\[2ex]
&+& e^{(2,4)}\,\omega_{A_{1}}\,\omega_{A_{2}}
+ e^{(2,5)}(\omega_{A_{1}}^2 \omega_{A_{2}} + \omega_{A_{1}}\,\omega_{A_{2}}^2)
+ e^{(2,6)} \,\omega_{A_{1}}^2 \omega_{A_{2}}^2\,.
\eea
\vspace{0.2cm}
The dependence of the quantities $e^{(i,j)}$ on the Symanzik
coefficients cannot be given in closed form; their values for the
Wilson and tree-level Symanzik gluon actions can be read in
Table~\ref{tab1}. For other actions the values of $e^{(i,j)}$ are
provided in electronic form (see Appendix~\ref{appC}).

\begin{table}[h]
\begin{center}
\begin{minipage}{13cm}
\begin{tabular}{lr@{}lr@{}l}
\hline
\hline
\multicolumn{1}{c}{$e^{(i,j)}$}&
\multicolumn{2}{c}{$\quad$Wilson}&
\multicolumn{2}{c}{tree-level Symanzik}\\
\hline
\hline
$\,\,e^{(1,1)}\,\,$    &\hspace*{0.5cm}-&9.83170$\,$   &\hspace*{0.75cm}-&7.21363\\
$\,\,e^{(1,2)}\,\,$    &\hspace*{0.5cm}&167.367$\,$   &\hspace*{0.75cm}&124.515\\
$\,\,e^{(1,3)}\,\,$    &\hspace*{0.5cm}-&710.612$\,$   &\hspace*{0.75cm}-&518.433\\
$\,\,e^{(1,4)}\,\,$    &\hspace*{0.5cm}-&2842.45$\,$   &\hspace*{0.75cm}-&2073.73\\
$\,\,e^{(1,5)}\,\,$    &\hspace*{0.5cm}&13134.2$\,$   &\hspace*{0.75cm}&9435.35\\
$\,\,e^{(1,6)}\,\,$    &\hspace*{0.5cm}-&64757.6$\,$   &\hspace*{0.75cm}-&45903.1\\
$\,\,e^{(2,1)}\,\,$    &\hspace*{0.5cm}&33.3933$\,$   &\hspace*{0.75cm}&27.1081\\
$\,\,e^{(2,2)}\,\,$    &\hspace*{0.5cm}-&342.525$\,$   &\hspace*{0.75cm}-&264.695\\
$\,\,e^{(2,3)}\,\,$    &\hspace*{0.5cm}&1174.37$\,$   &\hspace*{0.75cm}&885.215\\
$\,\,e^{(2,4)}\,\,$    &\hspace*{0.5cm}&4697.49$\,$   &\hspace*{0.75cm}&3540.86\\
$\,\,e^{(2,5)}\,\,$    &\hspace*{0.5cm}-&18790.0$\,$   &\hspace*{0.75cm}-&13960.0\\
$\,\,e^{(2,6)}\,\,$    &\hspace*{0.5cm}&82920.9$\,$   &\hspace*{0.75cm}&60910.8\\
\hline
\hline
\end{tabular}
\end{minipage}
\end{center}
\caption{Numerical coefficients $e^{(i,j)}$ of the propagator for the
Wilson and tree-level Symanzik actions.}
\label{tab1}
\end{table}

\newpage
\section{NOTATION IN MATHEMATICA FILE STAGGERED.M}
\label{appC}

The full body of our results can be accessed online through the
file Staggered.m, which is a Mathematica input file [?]. ???In this 
distribution package we include a collection of our
results in order to make them easily accessible to the
reader. More details and definitions can be found in Appendix C of
the paper. 
It includes the following.

\vspace*{0.7cm}
\noindent 
\subsection{One- and two-gluon doubly stout links}
The expressions for the one- and two-gluon ``doubly stout" links for
different stout parameters in the first and second smearing step:
\bea
\tilde{\tilde{U}}^{(1)}_{\mu}(x;{\omega_1,\omega_2}) &=& U1[x,{\rm mu},{\rm omega1},{\rm omega2}] \\
\tilde{\tilde{U}}^{(2)}_{\mu}(x;{\omega_1,\omega_2}) &=& U2[x,{\rm mu},{\rm omega1},{\rm omega2}] 
\eea
The arguments of U1 and U2 are the following:
\begin{itemize}
\item $x$: position of the link in the lattice
\item $\rm mu$: direction of the link
\item $\rm omega1$: the first
 stout parameter
\item $\rm omega2$: the second stout parameter
\end{itemize}
Moreover, the gluon field is denoted as\footnote{Regarding the
  ordering in products of gluon fields in $\tilde{\tilde{U}}^{(2)}_{\mu}$
see Note in Appendix~\ref{appA}.} $A_\sigma(x + a \hat{\tau}) \equiv$ A[sigma,x+tau].
The indices rho[1]-rho[4] appearing in $\tilde{\tilde{U}}^{(1)}_{\mu}$
and $\tilde{\tilde{U}}^{(2)}_{\mu}$ are dummy: a summation 
$\displaystyle\sum_{{\rm rho}[1]=+1}^{+4}$ is implied, but 
only in terms which contain rho[1]; similarly for rho[2]-rho[4].

\vspace*{0.7cm}
\noindent
\subsection{One-loop inverse propagator}
The one-loop inverse propagator is
\be
S^{-1}_{1-loop} = {\rm propagator[Action,alpha,omegaA1,omegaA2,g2tilde,m]} \,.
\label{SS}
\ee
This expression depends on the following variables.
\begin{itemize}
\item Action (selection of improved gauge action as follows~\cite{Alexandrou:2012mt}: 
\begin{itemize}
\item[1:] Plaquette, 
\item[2:] Tree Level Symanzik,
\item[3:] TILW ($\beta\,c_0=8.60$),
\item[4:] TILW ($\beta\,c_0=8.45$),
\item[5:] TILW ($\beta\,c_0=8.30$),
\item[6:] TILW ($\beta\,c_0=8.20$),
\item[7:] TILW ($\beta\,c_0=8.10$),
\item[8:] TILW ($\beta\,c_0=8.00$),
\item[9:] Iwasaki,
\item[10:] DBW2,
\end{itemize}
where $\displaystyle\beta=2N_c/g_0^2$.
\item alpha: gauge parameter (Landau/Feynman/generic correspond to 0/1/alpha)
\item $\rm omegaA1$($\rm omegaA2$): the first (second) stout
parameter coming from the smearing of the links appearing in the action
\item g2tilde: $\tilde{g}^2 \equiv \displaystyle\ggcf$ (g: coupling constant)
\item m: Lagrangian mass
\item pslash $\equiv \pslash$
\item p2 $\equiv \,p^2$
\end{itemize}

\vspace*{0.7cm}
\noindent
\subsection{Amputated Green's functions}
The amputated Green's functions relevant to the ultralocal operators:
\bea
\label{LS}
\Lambda_S^{1-loop} &=& {\rm \phantom{scalar}scalar}[{\rm Action,alpha,omegaA1,omegaA2,omegaO1,omegaO2,g2tilde,m}]   \\
\Lambda_P^{1-loop} &=& {\rm pseudoscalar}[{\rm Action,alpha,omegaA1,omegaA2,omegaO1,omegaO2,g2tilde,m}]   \\
\Lambda_V^{1-loop} &=& {\rm \phantom{scalar}vector}[{\rm Action,alpha,omegaA1,omegaA2,omegaO1,omegaO2,g2tilde,m}]   \\
\Lambda_A^{1-loop} &=& {\rm \phantom{oscalar}axial}[{\rm Action,alpha,omegaA1,omegaA2,omegaO1,omegaO2,g2tilde,m}]   \\
\Lambda_T^{1-loop} &=& {\rm \phantom{scalar}tensor}[{\rm Action,alpha,omegaA1,omegaA2,omegaO1,omegaO2,g2tilde,m}]  \,.
\label{LT}
\eea
where
\begin{itemize}
\item $\rm omegaO1$($\rm omegaO2$): the first (second) stout
parameter coming from the smearing of the links appearing in the
bilinears
\item p[mu] $\equiv p_\mu$
\item p[nu] $\equiv \,p_\nu$
\item gamma5 $\equiv\,\gamma_5$
\item gamma1 $\equiv \gamma_\mu$
\item gamma2 $\equiv \gamma_\nu$
\item gamma5gamma1 $\equiv\,\gamma_5\,\gamma_\mu$
\item gamma5pslash $\equiv\,\gamma_5\,\pslash$
\item gamma5gamma1pslash $\equiv\,\gamma_5\,\gamma_\mu\,\pslash$
\item gamma1gamma2 $\equiv \gamma_\mu\,\gamma_\nu$
\item gamma2gamma1 $\equiv \gamma_\nu\,\gamma_\mu$
\item gamma1gamma2pslash $\equiv \gamma_\mu\,\gamma_\nu\,\pslash$
\item gamma1pslash $\equiv\,\gamma_\mu\,\pslash$
\item gamma2pslash $\equiv\,\gamma_\nu\,\pslash$
\end{itemize}
We note that Eqs. (\ref{LS}) $-$ (\ref{LT}) hold for fermions with the
same Lagrangian mass. 

\vspace*{0.7cm}
\noindent
\subsection{RFs}
The RFs of the fermion field and fermion
bilinears in the RI$'$-MOM scheme:
\bea
\Zq^{\rm RI'} &=& {\rm zq[Action,alpha,omegaA1,omegaA2,g2tilde,p2]} \\
\ZS^{\rm RI'} &=& {\rm zs[Action,alpha,omegaA1,omegaA2,omegaO1,omegaO2,g2tilde,p2]} \\
\ZP^{\rm RI'} &=& {\rm zp[Action,alpha,omegaA1,omegaA2,omegaO1,omegaO2,g2tilde,p2]} \\
\ZV^{\rm RI'} &=& {\rm zv[Action,alpha,omegaA1,omegaA2,omegaO1,omegaO2,g2tilde,p2]} \\
\ZA^{\rm RI'} &=& {\rm za[Action,alpha,omegaA1,omegaA2,omegaO1,omegaO2,g2tilde,p2]} \\
\ZT^{\rm RI'} &=& {\rm zt[Action,alpha,omegaA1,omegaA2,omegaO1,omegaO2,g2tilde,p2]}\,. 
\label{ZTRI}
\eea
For convenience, all quantities in Eqs.~(\ref{SS}) $-$ (\ref{ZTRI}) may
be also retrieved using only their first argument; thus, for example:
vector[9] will result in $\Lambda_V^{1-loop}$ for the Iwasaki action
with generic values of $\alpha,\, \omega_{A_i},\,\omega_{{\cal O}_i},\,g,\,m,\,p$.

%%%%%%%%%%%%%%%%%%%%%%%%%%%%%%%%%%%%%%%%%%%%%%%%%%%%%%%%%%%%%%%%%%%%
% ========================= REFERENCES =========================

\bibliographystyle{apsrev}                     % Style for bibliography
\bibliography{Staggered_VAP}

\begin{thebibliography}{35}
\expandafter\ifx\csname natexlab\endcsname\relax\def\natexlab#1{#1}\fi
\expandafter\ifx\csname bibnamefont\endcsname\relax
  \def\bibnamefont#1{#1}\fi
\expandafter\ifx\csname bibfnamefont\endcsname\relax
  \def\bibfnamefont#1{#1}\fi
\expandafter\ifx\csname citenamefont\endcsname\relax
  \def\citenamefont#1{#1}\fi
\expandafter\ifx\csname url\endcsname\relax
  \def\url#1{\texttt{#1}}\fi
\expandafter\ifx\csname urlprefix\endcsname\relax\def\urlprefix{URL }\fi
\providecommand{\bibinfo}[2]{#2}
\providecommand{\eprint}[2][]{\url{#2}}

\bibitem[{\citenamefont{H$\ddot{\rm a}$gler}(2010)}]{Hagler:2009ni}
\bibinfo{author}{\bibfnamefont{P.}~\bibnamefont{H$\ddot{\rm a}$gler}},
  \bibinfo{journal}{Phys. Rept.} \textbf{\bibinfo{volume}{490}},
  \bibinfo{pages}{49} (\bibinfo{year}{2010}), \eprint{[arXiv:0912.5483]}.

\bibitem[{\citenamefont{Doi}(2012)}]{Doi:2012ab}
\bibinfo{author}{\bibfnamefont{T.}~\bibnamefont{Doi}}
  (\bibinfo{collaboration}{HAL QCD Collaboration}), \bibinfo{journal}{PoS}
  \textbf{\bibinfo{volume}{LATTICE2012}}, \bibinfo{pages}{009}
  (\bibinfo{year}{2012}), \eprint{[arXiv:1212.1572]}.

\bibitem[{\citenamefont{Lin}(2012)}]{Lin:2012ev}
\bibinfo{author}{\bibfnamefont{H.-W.} \bibnamefont{Lin}},
  \bibinfo{journal}{PoS} \textbf{\bibinfo{volume}{LATTICE2012}},
  \bibinfo{pages}{013} (\bibinfo{year}{2012}), \eprint{[arXiv:1212.6849]}.

\bibitem[{\citenamefont{Kogut and Susskind}(1975)}]{Kogut:1974ag}
\bibinfo{author}{\bibfnamefont{J.~B.} \bibnamefont{Kogut}} \bibnamefont{and}
  \bibinfo{author}{\bibfnamefont{L.}~\bibnamefont{Susskind}},
  \bibinfo{journal}{Phys. Rev.} \textbf{\bibinfo{volume}{D11}},
  \bibinfo{pages}{395} (\bibinfo{year}{1975}).

\bibitem[{\citenamefont{Bazavov et~al.}(2013)\citenamefont{Bazavov, Bernard,
  DeTar, Freeman, Gottlieb et~al.}}]{Bazavov:2012zad}
\bibinfo{author}{\bibfnamefont{A.}~\bibnamefont{Bazavov}},
  \bibinfo{author}{\bibfnamefont{C.}~\bibnamefont{Bernard}},
  \bibinfo{author}{\bibfnamefont{C.}~\bibnamefont{DeTar}},
  \bibinfo{author}{\bibfnamefont{W.}~\bibnamefont{Freeman}},
  \bibinfo{author}{\bibfnamefont{S.}~\bibnamefont{Gottlieb}},
  \bibnamefont{et~al.} (\bibinfo{collaboration}{MILC Collaboration}),
  \bibinfo{journal}{Phys. Rev.} \textbf{\bibinfo{volume}{D87}},
  \bibinfo{pages}{054503} (\bibinfo{year}{2013}), \eprint{[arXiv:1212.4768]}.

\bibitem[{\citenamefont{Allison et~al.}(2008)\citenamefont{Allison, Dalgic,
  Davies, Follana, Horgan et~al.}}]{Allison:2008xk}
\bibinfo{author}{\bibfnamefont{I.}~\bibnamefont{Allison}},
  \bibinfo{author}{\bibfnamefont{E.}~\bibnamefont{Dalgic}},
  \bibinfo{author}{\bibfnamefont{C.}~\bibnamefont{Davies}},
  \bibinfo{author}{\bibfnamefont{E.}~\bibnamefont{Follana}},
  \bibinfo{author}{\bibfnamefont{R.}~\bibnamefont{Horgan}},
  \bibnamefont{et~al.} (\bibinfo{collaboration}{HPQCD Collaboration}),
  \bibinfo{journal}{Phys. Rev.} \textbf{\bibinfo{volume}{D78}},
  \bibinfo{pages}{054513} (\bibinfo{year}{2008}), \eprint{[arXiv:0805.2999]}.

\bibitem[{\citenamefont{Bali et~al.}(2012{\natexlab{a}})\citenamefont{Bali,
  Bruckmann, Endr\H{o}di, Fodor, Katz et~al.}}]{Bali:2011qj}
\bibinfo{author}{\bibfnamefont{G.}~\bibnamefont{Bali}},
  \bibinfo{author}{\bibfnamefont{F.}~\bibnamefont{Bruckmann}},
  \bibinfo{author}{\bibfnamefont{G.}~\bibnamefont{Endr\H{o}di}},
  \bibinfo{author}{\bibfnamefont{Z.}~\bibnamefont{Fodor}},
  \bibinfo{author}{\bibfnamefont{S.}~\bibnamefont{Katz}}, \bibnamefont{et~al.},
  \bibinfo{journal}{JHEP} \textbf{\bibinfo{volume}{1202}}, \bibinfo{pages}{044}
  (\bibinfo{year}{2012}{\natexlab{a}}), \eprint{[arXiv:1111.4956]}.

\bibitem[{\citenamefont{Bali et~al.}(2012{\natexlab{b}})\citenamefont{Bali,
  Bruckmann, Constantinou, Costa, Endr\H{o}di et~al.}}]{Bali:2012jv}
\bibinfo{author}{\bibfnamefont{G.}~\bibnamefont{Bali}},
  \bibinfo{author}{\bibfnamefont{F.}~\bibnamefont{Bruckmann}},
  \bibinfo{author}{\bibfnamefont{M.}~\bibnamefont{Constantinou}},
  \bibinfo{author}{\bibfnamefont{M.}~\bibnamefont{Costa}},
  \bibinfo{author}{\bibfnamefont{G.}~\bibnamefont{Endr\H{o}di}},
  \bibnamefont{et~al.}, \bibinfo{journal}{Phys. Rev.}
  \textbf{\bibinfo{volume}{D86}}, \bibinfo{pages}{094512}
  (\bibinfo{year}{2012}{\natexlab{b}}), \eprint{[arXiv:1209.6015]}.

\bibitem[{\citenamefont{Aoki et~al.}(2006)\citenamefont{Aoki, Fodor, Katz, and
  Szabo}}]{Aoki:2005vt}
\bibinfo{author}{\bibfnamefont{Y.}~\bibnamefont{Aoki}},
  \bibinfo{author}{\bibfnamefont{Z.}~\bibnamefont{Fodor}},
  \bibinfo{author}{\bibfnamefont{S.~D.} \bibnamefont{Katz}}, \bibnamefont{and}
  \bibinfo{author}{\bibfnamefont{K.}~\bibnamefont{Szabo}},
  \bibinfo{journal}{JHEP} \textbf{\bibinfo{volume}{0601}}, \bibinfo{pages}{089}
  (\bibinfo{year}{2006}), \eprint{[hep-lat/0510084]}.

\bibitem[{\citenamefont{Bors\'anyi
  et~al.}(2011{\natexlab{a}})\citenamefont{Bors\'anyi, Fodor, Katz, Krieg,
  Ratti et~al.}}]{Borsanyi:2011bm}
\bibinfo{author}{\bibfnamefont{S.}~\bibnamefont{Bors\'anyi}},
  \bibinfo{author}{\bibfnamefont{Z.}~\bibnamefont{Fodor}},
  \bibinfo{author}{\bibfnamefont{S.}~\bibnamefont{Katz}},
  \bibinfo{author}{\bibfnamefont{S.}~\bibnamefont{Krieg}},
  \bibinfo{author}{\bibfnamefont{C.}~\bibnamefont{Ratti}}, \bibnamefont{et~al.}
  (\bibinfo{collaboration}{Wuppertal-Budapest Collaboration}),
  \bibinfo{journal}{J. Phys.} \textbf{\bibinfo{volume}{G38}},
  \bibinfo{pages}{124060} (\bibinfo{year}{2011}{\natexlab{a}}),
  \eprint{[arXiv:1109.5030]}.

\bibitem[{\citenamefont{Bazavov et~al.}(2012)\citenamefont{Bazavov,
  Bhattacharya, Cheng, DeTar, Ding et~al.}}]{Bazavov:2011nk}
\bibinfo{author}{\bibfnamefont{A.}~\bibnamefont{Bazavov}},
  \bibinfo{author}{\bibfnamefont{T.}~\bibnamefont{Bhattacharya}},
  \bibinfo{author}{\bibfnamefont{M.}~\bibnamefont{Cheng}},
  \bibinfo{author}{\bibfnamefont{C.}~\bibnamefont{DeTar}},
  \bibinfo{author}{\bibfnamefont{H.}~\bibnamefont{Ding}}, \bibnamefont{et~al.},
  \bibinfo{journal}{Phys. Rev.} \textbf{\bibinfo{volume}{D85}},
  \bibinfo{pages}{054503} (\bibinfo{year}{2012}), \eprint{[arXiv:1111.1710]}.

\bibitem[{\citenamefont{Capitani}(2003)}]{Capitani:2002mp}
\bibinfo{author}{\bibfnamefont{S.}~\bibnamefont{Capitani}},
  \bibinfo{journal}{Phys. Rept.} \textbf{\bibinfo{volume}{382}},
  \bibinfo{pages}{113} (\bibinfo{year}{2003}), \eprint{[hep-lat/0211036]}.

\bibitem[{\citenamefont{Patel and Sharpe}(1993)}]{Patel:1992vu}
\bibinfo{author}{\bibfnamefont{A.}~\bibnamefont{Patel}} \bibnamefont{and}
  \bibinfo{author}{\bibfnamefont{S.~R.} \bibnamefont{Sharpe}},
  \bibinfo{journal}{Nucl. Phys.} \textbf{\bibinfo{volume}{B395}},
  \bibinfo{pages}{701} (\bibinfo{year}{1993}), \eprint{[hep-lat/9210039]}.

\bibitem[{\citenamefont{Aoki et~al.}(2003)\citenamefont{Aoki, Izubuchi,
  Kuramashi, and Taniguchi}}]{Aoki:2002iq}
\bibinfo{author}{\bibfnamefont{S.}~\bibnamefont{Aoki}},
  \bibinfo{author}{\bibfnamefont{T.}~\bibnamefont{Izubuchi}},
  \bibinfo{author}{\bibfnamefont{Y.}~\bibnamefont{Kuramashi}},
  \bibnamefont{and}
  \bibinfo{author}{\bibfnamefont{Y.}~\bibnamefont{Taniguchi}},
  \bibinfo{journal}{Phys. Rev.} \textbf{\bibinfo{volume}{D67}},
  \bibinfo{pages}{094502} (\bibinfo{year}{2003}), \eprint{[hep-lat/0206013]}.

\bibitem[{\citenamefont{Mason et~al.}(2006)\citenamefont{Mason, Trottier,
  Horgan, Davies, and Lepage}}]{Mason:2005bj}
\bibinfo{author}{\bibfnamefont{Q.}~\bibnamefont{Mason}},
  \bibinfo{author}{\bibfnamefont{H.~D.} \bibnamefont{Trottier}},
  \bibinfo{author}{\bibfnamefont{R.}~\bibnamefont{Horgan}},
  \bibinfo{author}{\bibfnamefont{C.~T.} \bibnamefont{Davies}},
  \bibnamefont{and} \bibinfo{author}{\bibfnamefont{G.~P.} \bibnamefont{Lepage}}
  (\bibinfo{collaboration}{HPQCD Collaboration}), \bibinfo{journal}{Phys. Rev.}
  \textbf{\bibinfo{volume}{D73}}, \bibinfo{pages}{114501}
  (\bibinfo{year}{2006}), \eprint{[hep-ph/0511160]}.

\bibitem[{\citenamefont{Skouroupathis and
  Panagopoulos}(2007)}]{Skouroupathis:2007jd}
\bibinfo{author}{\bibfnamefont{A.}~\bibnamefont{Skouroupathis}}
  \bibnamefont{and}
  \bibinfo{author}{\bibfnamefont{H.}~\bibnamefont{Panagopoulos}},
  \bibinfo{journal}{Phys. Rev.} \textbf{\bibinfo{volume}{D76}},
  \bibinfo{pages}{094514} (\bibinfo{year}{2007}), \eprint{[arXiv:0707.2906]}.

\bibitem[{\citenamefont{Alexandrou et~al.}(2011)\citenamefont{Alexandrou,
  Constantinou, Korzec, Panagopoulos, and Stylianou}}]{Alexandrou:2010me}
\bibinfo{author}{\bibfnamefont{C.}~\bibnamefont{Alexandrou}},
  \bibinfo{author}{\bibfnamefont{M.}~\bibnamefont{Constantinou}},
  \bibinfo{author}{\bibfnamefont{T.}~\bibnamefont{Korzec}},
  \bibinfo{author}{\bibfnamefont{H.}~\bibnamefont{Panagopoulos}},
  \bibnamefont{and}
  \bibinfo{author}{\bibfnamefont{F.}~\bibnamefont{Stylianou}},
  \bibinfo{journal}{Phys. Rev.} \textbf{\bibinfo{volume}{D83}},
  \bibinfo{pages}{014503} (\bibinfo{year}{2011}), \eprint{[arXiv:1006.1920]}.

\bibitem[{\citenamefont{Lee and Sharpe}(2002)}]{Lee:2002ui}
\bibinfo{author}{\bibfnamefont{W.}~\bibnamefont{Lee}} \bibnamefont{and}
  \bibinfo{author}{\bibfnamefont{S.~R.} \bibnamefont{Sharpe}},
  \bibinfo{journal}{Phys. Rev.} \textbf{\bibinfo{volume}{D66}},
  \bibinfo{pages}{114501} (\bibinfo{year}{2002}), \eprint{[hep-lat/0208018]}.

\bibitem[{\citenamefont{Kim et~al.}(2010)\citenamefont{Kim, Lee, and
  Sharpe}}]{Kim:2010fj}
\bibinfo{author}{\bibfnamefont{J.}~\bibnamefont{Kim}},
  \bibinfo{author}{\bibfnamefont{W.}~\bibnamefont{Lee}}, \bibnamefont{and}
  \bibinfo{author}{\bibfnamefont{S.~R.} \bibnamefont{Sharpe}},
  \bibinfo{journal}{Phys. Rev.} \textbf{\bibinfo{volume}{D81}},
  \bibinfo{pages}{114503} (\bibinfo{year}{2010}), \eprint{[arXiv:1004.4039]}.

\bibitem[{\citenamefont{Morningstar and Peardon}(2004)}]{Morningstar:2003gk}
\bibinfo{author}{\bibfnamefont{C.}~\bibnamefont{Morningstar}} \bibnamefont{and}
  \bibinfo{author}{\bibfnamefont{M.~J.} \bibnamefont{Peardon}},
  \bibinfo{journal}{Phys. Rev.} \textbf{\bibinfo{volume}{D69}},
  \bibinfo{pages}{054501} (\bibinfo{year}{2004}), \eprint{[hep-lat/0311018]}.

\bibitem[{\citenamefont{Skouroupathis and
  Panagopoulos}(2009)}]{Skouroupathis:2008mf}
\bibinfo{author}{\bibfnamefont{A.}~\bibnamefont{Skouroupathis}}
  \bibnamefont{and}
  \bibinfo{author}{\bibfnamefont{H.}~\bibnamefont{Panagopoulos}},
  \bibinfo{journal}{Phys. Rev.} \textbf{\bibinfo{volume}{D79}},
  \bibinfo{pages}{094508} (\bibinfo{year}{2009}), \eprint{[arXiv:0811.4264]}.

\bibitem[{\citenamefont{Capitani et~al.}(2001)\citenamefont{Capitani,
  G$\ddot{\rm o}$ckeler, Horsley, Perlt, Rakow et~al.}}]{Capitani:2000xi}
\bibinfo{author}{\bibfnamefont{S.}~\bibnamefont{Capitani}},
  \bibinfo{author}{\bibfnamefont{M.}~\bibnamefont{G$\ddot{\rm o}$ckeler}},
  \bibinfo{author}{\bibfnamefont{R.}~\bibnamefont{Horsley}},
  \bibinfo{author}{\bibfnamefont{H.}~\bibnamefont{Perlt}},
  \bibinfo{author}{\bibfnamefont{P.~E.~L.} \bibnamefont{Rakow}},
  \bibnamefont{et~al.}, \bibinfo{journal}{Nucl. Phys.}
  \textbf{\bibinfo{volume}{B593}}, \bibinfo{pages}{183} (\bibinfo{year}{2001}),
  \eprint{[hep-lat/0007004]}.

\bibitem[{\citenamefont{Constantinou et~al.}(2009)\citenamefont{Constantinou,
  Lubicz, Panagopoulos, and Stylianou}}]{Constantinou:2009tr}
\bibinfo{author}{\bibfnamefont{M.}~\bibnamefont{Constantinou}},
  \bibinfo{author}{\bibfnamefont{V.}~\bibnamefont{Lubicz}},
  \bibinfo{author}{\bibfnamefont{H.}~\bibnamefont{Panagopoulos}},
  \bibnamefont{and}
  \bibinfo{author}{\bibfnamefont{F.}~\bibnamefont{Stylianou}},
  \bibinfo{journal}{JHEP} \textbf{\bibinfo{volume}{0910}}, \bibinfo{pages}{064}
  (\bibinfo{year}{2009}), \eprint{[arXiv:0907.0381]}.

\bibitem[{\citenamefont{Alexandrou et~al.}(2012)\citenamefont{Alexandrou,
  Constantinou, Korzec, Panagopoulos, and Stylianou}}]{Alexandrou:2012mt}
\bibinfo{author}{\bibfnamefont{C.}~\bibnamefont{Alexandrou}},
  \bibinfo{author}{\bibfnamefont{M.}~\bibnamefont{Constantinou}},
  \bibinfo{author}{\bibfnamefont{T.}~\bibnamefont{Korzec}},
  \bibinfo{author}{\bibfnamefont{H.}~\bibnamefont{Panagopoulos}},
  \bibnamefont{and}
  \bibinfo{author}{\bibfnamefont{F.}~\bibnamefont{Stylianou}},
  \bibinfo{journal}{Phys. Rev.} \textbf{\bibinfo{volume}{D86}},
  \bibinfo{pages}{014505} (\bibinfo{year}{2012}), \eprint{[arXiv:1201.5025]}.

\bibitem[{\citenamefont{Horsley et~al.}(2008)\citenamefont{Horsley, Perlt,
  Rakow, Schierholz, and Schiller}}]{Horsley:2008ap}
\bibinfo{author}{\bibfnamefont{R.}~\bibnamefont{Horsley}},
  \bibinfo{author}{\bibfnamefont{H.}~\bibnamefont{Perlt}},
  \bibinfo{author}{\bibfnamefont{P.~E.~L.} \bibnamefont{Rakow}},
  \bibinfo{author}{\bibfnamefont{G.}~\bibnamefont{Schierholz}},
  \bibnamefont{and} \bibinfo{author}{\bibfnamefont{A.}~\bibnamefont{Schiller}},
  \bibinfo{journal}{Phys. Rev.} \textbf{\bibinfo{volume}{D78}},
  \bibinfo{pages}{054504} (\bibinfo{year}{2008}), \eprint{[arXiv:0807.0345]}.

\bibitem[{\citenamefont{Daniel and Sheard}(1988)}]{Daniel:1987aa}
\bibinfo{author}{\bibfnamefont{D.}~\bibnamefont{Daniel}} \bibnamefont{and}
  \bibinfo{author}{\bibfnamefont{S.~N.} \bibnamefont{Sheard}},
  \bibinfo{journal}{Nucl. Phys.} \textbf{\bibinfo{volume}{B302}},
  \bibinfo{pages}{471} (\bibinfo{year}{1988}).

\bibitem[{\citenamefont{Ishizuka and Shizawa}(1994)}]{Ishizuka:1993fs}
\bibinfo{author}{\bibfnamefont{N.}~\bibnamefont{Ishizuka}} \bibnamefont{and}
  \bibinfo{author}{\bibfnamefont{Y.}~\bibnamefont{Shizawa}},
  \bibinfo{journal}{Phys. Rev.} \textbf{\bibinfo{volume}{D49}},
  \bibinfo{pages}{3519} (\bibinfo{year}{1994}), \eprint{[hep-lat/9308008]}.

\bibitem[{\citenamefont{Gracey}(2003)}]{Gracey:2003yr}
\bibinfo{author}{\bibfnamefont{J.}~\bibnamefont{Gracey}},
  \bibinfo{journal}{Nucl. Phys.} \textbf{\bibinfo{volume}{B662}},
  \bibinfo{pages}{247} (\bibinfo{year}{2003}), \eprint{[hep-ph/0304113]}.

\bibitem[{\citenamefont{Buras and Weisz}(1990)}]{Buras:1989xd}
\bibinfo{author}{\bibfnamefont{A.~J.} \bibnamefont{Buras}} \bibnamefont{and}
  \bibinfo{author}{\bibfnamefont{P.~H.} \bibnamefont{Weisz}},
  \bibinfo{journal}{Nucl. Phys.} \textbf{\bibinfo{volume}{B333}},
  \bibinfo{pages}{66} (\bibinfo{year}{1990}).

\bibitem[{\citenamefont{'t~Hooft and Veltman}(1972)}]{'tHooft:1972fi}
\bibinfo{author}{\bibfnamefont{G.}~\bibnamefont{'t~Hooft}} \bibnamefont{and}
  \bibinfo{author}{\bibfnamefont{M.}~\bibnamefont{Veltman}},
  \bibinfo{journal}{Nucl. Phys.} \textbf{\bibinfo{volume}{B44}},
  \bibinfo{pages}{189} (\bibinfo{year}{1972}).

\bibitem[{\citenamefont{Larin}(1993)}]{Larin:1993tq}
\bibinfo{author}{\bibfnamefont{S.}~\bibnamefont{Larin}},
  \bibinfo{journal}{Phys. Lett.} \textbf{\bibinfo{volume}{B303}},
  \bibinfo{pages}{113} (\bibinfo{year}{1993}), \eprint{[hep-ph/9302240,
  containing an extra section]}.

\bibitem[{\citenamefont{Kim et~al.}(2011)\citenamefont{Kim, Lee, and
  Sharpe}}]{Kim:2011pz}
\bibinfo{author}{\bibfnamefont{J.}~\bibnamefont{Kim}},
  \bibinfo{author}{\bibfnamefont{W.}~\bibnamefont{Lee}}, \bibnamefont{and}
  \bibinfo{author}{\bibfnamefont{S.~R.} \bibnamefont{Sharpe}},
  \bibinfo{journal}{Phys. Rev.} \textbf{\bibinfo{volume}{D83}},
  \bibinfo{pages}{094503} (\bibinfo{year}{2011}), \eprint{[arXiv:1102.1774]}.

\bibitem[{\citenamefont{Capitani et~al.}(2006)\citenamefont{Capitani, Durr, and
  Hoelbling}}]{Capitani:2006ni}
\bibinfo{author}{\bibfnamefont{S.}~\bibnamefont{Capitani}},
  \bibinfo{author}{\bibfnamefont{S.}~\bibnamefont{Durr}}, \bibnamefont{and}
  \bibinfo{author}{\bibfnamefont{C.}~\bibnamefont{Hoelbling}},
  \bibinfo{journal}{JHEP} \textbf{\bibinfo{volume}{0611}}, \bibinfo{pages}{028}
  (\bibinfo{year}{2006}), \eprint{[hep-lat/0607006]}.

\bibitem[{\citenamefont{Bors\'anyi
  et~al.}(2011{\natexlab{b}})\citenamefont{Bors\'anyi, Fodor, Hoelbling, Katz,
  Krieg et~al.}}]{Borsanyi:2011kg}
\bibinfo{author}{\bibfnamefont{S.}~\bibnamefont{Bors\'anyi}},
  \bibinfo{author}{\bibfnamefont{Z.}~\bibnamefont{Fodor}},
  \bibinfo{author}{\bibfnamefont{C.}~\bibnamefont{Hoelbling}},
  \bibinfo{author}{\bibfnamefont{S.~D.} \bibnamefont{Katz}},
  \bibinfo{author}{\bibfnamefont{S.}~\bibnamefont{Krieg}},
  \bibnamefont{et~al.}, \bibinfo{journal}{PoS}
  \textbf{\bibinfo{volume}{LATTICE2011}}, \bibinfo{pages}{209}
  (\bibinfo{year}{2011}{\natexlab{b}}), \eprint{[arXiv:1111.3500]}.

\bibitem[{\citenamefont{Constantinou et~al.}(2013)\citenamefont{Constantinou,
  Costa, G$\ddot{\rm o}$ckeler, Horsley, Panagopoulos
  et~al.}}]{Constantinou:2013ada}
\bibinfo{author}{\bibfnamefont{M.}~\bibnamefont{Constantinou}},
  \bibinfo{author}{\bibfnamefont{M.}~\bibnamefont{Costa}},
  \bibinfo{author}{\bibfnamefont{M.}~\bibnamefont{G$\ddot{\rm o}$ckeler}},
  \bibinfo{author}{\bibfnamefont{R.}~\bibnamefont{Horsley}},
  \bibinfo{author}{\bibfnamefont{H.}~\bibnamefont{Panagopoulos}},
  \bibnamefont{et~al.}, \bibinfo{journal}{Phys. Rev.}
  \textbf{\bibinfo{volume}{D87}}, \bibinfo{pages}{096019}
  (\bibinfo{year}{2013}), \eprint{[arXiv:1303.6776]}.

\end{thebibliography}

\end{document}